\documentclass{article}

\usepackage{arxiv}

\usepackage{soul}
\usepackage[utf8]{inputenc} 
\usepackage[T1]{fontenc}    
\usepackage{xr}
\usepackage{hyperref}       
\usepackage{url}            
\usepackage{booktabs}       
\usepackage{amsfonts}       
\usepackage{nicefrac}       
\usepackage{microtype}      
\usepackage{lipsum}
\usepackage{graphicx}
\graphicspath{ {./images/} }
\usepackage{siunitx}
\usepackage{amsmath}
\usepackage[scientific-notation=true]{siunitx}
\usepackage{bbm}
\usepackage{caption}
\usepackage{algorithm}
\usepackage{xcolor}
\usepackage{multirow}
\usepackage{algpseudocode}
\usepackage{subcaption}
\usepackage[superscript,biblabel]{cite}
\usepackage{tikz}
\usepackage{verbatim}
\usepackage[export]{adjustbox}

\definecolor{green1}{HTML}{82B776}
\definecolor{blue1}{HTML}{3F96F5}
\definecolor{orange1}{HTML}{DA6412}
\definecolor{purple1}{HTML}{7570B3}

\newcommand{\ic}{$\text{IC}_{\text{50 }}$}
\title{Identifying Bayesian Optimal Experiments for Uncertain Biochemical Pathway Models} 

\author{
Natalie M. Isenberg\\
  Brookhaven National Laboratory\\
  Upton NY, 11973, USA \\
  \texttt{nisenberg@bnl.gov} \\
   \And
 Susan D. Mertins \\
 Fredrick National Laboratory for Cancer Research\\
 Fredrick MD, 21702, USA\\
  \texttt{susan.mertins2@nih.gov} \\
  \And
 Byung-Jun Yoon \\
  Texas A\&M University\\
  College Station TX, 77843, USA\\
  \texttt{bjyoon@tamu.edu} \\
    \And
    Kristofer Reyes \\
  University at Buffalo\\
  Buffalo NY, 14260, USA\\
  \texttt{kreyes3@buffalo.edu} \\
    \And
 Nathan M. Urban \\
  Brookhaven National Laboratory\\
  Upton NY, 11973, USA \\
  \texttt{nurban@bnl.gov}
}

\begin{document}
\maketitle

\section*{Summary}
Pharmacodynamic (PD) models are mathematical models of cellular reaction networks that include drug mechanisms of action. These models are useful for studying predictive therapeutic outcomes of novel drug therapies \textit{in silico}. However, PD models are known to possess significant uncertainty with respect to constituent parameter data, leading to uncertainty in the model predictions. Furthermore, experimental data to calibrate these models is often limited or unavailable for novel pathways. In this study, we present a Bayesian optimal experimental design approach for improving PD model prediction accuracy. We then apply our method using simulated experimental data to account for uncertainty in hypothetical laboratory measurements. This leads to a probabilistic prediction of drug performance and a quantitative measure of which prospective laboratory experiment will optimally reduce prediction uncertainty in the PD model. The methods proposed here provide a way forward for uncertainty quantification and guided experimental design for models of novel biological pathways.


\section{Introduction}\label{sec:intro}

The intricate and lengthy process of discovering novel drugs stands to gain significant advantages through the application of computational methods.\cite{sabe2021current} These methods provide efficient and cost-effective avenues for exploring expansive chemical spaces, predicting molecular properties, and optimizing potential drug candidates.
Dynamic models of biological systems (e.g., pharmacokinetic (PK) or pharmacodynamic (PD) models) are useful tools for studying biomolecular processes. 
Such models have been applied to study inter- and intra-cellular phenomena, including regulatory, metabolic, and signalling processes within human cells.\cite{breitling2008structured} \cite{chen2010classic} \cite{nijhout2019systems}
These models are derived from first-principles approximations of the complex dynamics that occur \textit{in vivo}. 
The utility of these biological models lies in their simplicity; tractably capturing qualitative system behaviors at the expense of prediction accuracy. 
Although these dynamic models are often used in preclinical drug development to determine \textit{in vivo} drug response\cite{braakman2022evaluation}, they are rarely applied in drug discovery. 
Biological pharmacodynamic models are valuable tools in both the optimal design and validation phases of identifying novel drug candidates, making them crucial for computational drug discovery. 
Furthermore, regulatory agencies are beginning to accept \textit{in silico} studies as part of the validation and testing of new medical therapies and technologies.\cite{viceconti2021silico}\cite{pappalardo2019silico}
This underscores a need to formalise approaches for uncertainty quantification and model improvement for pharmacodynamic models. 

Uncertainty quantification is a critical analysis that is often overlooked in computational drug design, despite its relevance and available software tools\cite{mitra2019parameter}. 
Previous studies have applied methods of uncertainty quantification to models for calibrating algorithmic parameters in automated diabetes treatment systems\cite{reenberg2022high}, a PK/PD cancer model for the antivascular endothelial growth factor in the present of a cancer therapeutic agent\cite{wu2019hierarchical}, the calibration of dose-response relationships for the $\gamma$-H2AX assay to predict radiation exposure\cite{einbeck2018statistical}, and peptide-MHC affinity prediction using a data-driven model\cite{zeng2019quantification}.
In the aforementioned studies, experimental or clinical data was readily available for the proposed uncertainty quantification procedures. 
For many PD models, the scarcity of quantitative data for calibrating these models has led to significant challenges in uncertainty quantification. 
This limited availability of experimental data results in non-unique and/or unconstrained parameter estimations, leading to issues of nonidentifiability \cite{raue2009structural}. 
Researchers have acknowledged the presence of significant parameter correlations, nonidentifiability, and parameter sloppiness in dynamic biology models, where parameters can vary over orders of magnitude without significantly affecting model output \cite{CHIS2016147}. 
Given these limitations, it is evident that incorporating uncertainty quantification to biological dynamic models, even in the absence of available data, is crucial if they are to be used for computational drug discovery and evaluation.

Optimal experimental design (OED) has been proposed as a method for improving model parameter estimates in biological pharmacodynamic models. In OED, existing measured data is used to determine what new experiment(s) should be done to best reduce parameter uncertainties \cite{raue2010identifiability}. 
Several works have applied different flavors of optimal experimental design to dynamic biological models. 
Work done by Liepe et al.\cite{liepe2013maximizing} demonstrates OED to maximize expected information gained in a signaling pathway model featuring 6 differential equations, 4 uncertain parameters, and selecting between 5 experimental outcomes.
Researchers in Bandara et al.\cite{bandara2009optimal} showed that a sequential experimental design for a cell signaling model, coupled with fluorescence microscopy-based experimental data, was able to reduce uncertainty in some parameters, while it increased in others due to nonidentifiability. 
This work also provides an example of selecting optimal experiments based on an experimentally relevant metric. This approach is in opposition to the more common purely information-theoretic approach. We note that the model used in this study is relatively small, only featuring 4 differential equations, 4 uncertain rate constants, and selecting between 3 experimental possibilities. 
Researchers in Eriksson et al.\cite{eriksson2019uncertainty} utilized approximate Bayesian computation to identify approximate posterior distributions for uncertain model parameters. 
This work presents a larger scale problem, with 34 reactions for 25 species, 34 uncertain parameters, and 6 experimental possibilities from which to select. 
This approximate approach is applicable to scenarios where the the data-generation distribution (likelihood function) is not available in analytic form. 

In this paper, we utilize methods from Bayesian optimal experimental design using exact Bayesian inference and Hamiltonian Monte Carlo (HMC) to recommend experiments for mechanistic model improvement. 
Our focus is on a dynamic model of programmed cell death, i.e., \textit{apoptosis}.
This model predicts synthetic lethality in cancer in the presence of a \textit{PARP1} inhibitor. 
Therefore, when we refer to selecting an optimal \textit{experimental design}, we mean ``which species should we measure from a \textit{PARP1}-inhibited cell experiment to improve confidence in simulated lethality predictions of a given inhibitor?''
Our goal is to first understand the impact of parameter uncertainty on the predictive reliability of this model. 
Then, the aim is to recommend experimental measurements that can be done in the lab to maximize confidence in simulated lethality predictions. To do this, we conduct parameter inference using HMC for large volumes of data generated from simulated experiments. This way, we can determine which experimental data, in expectation, would most reduce uncertainty in modeled drug performance predictions. 
Such an approach provides researchers with a quantitative method for linking physical experiment design to model improvement in the absence of existing data.

It is important to note that while our optimal experimental design study is conducted with simulated experimental data, and the results are thus subject to our modeling assumptions, the dynamic model underpinning our analysis contains biologically relevant and measurable parameters, thus offering insights that can guide future experimental endeavors.
Furthermore, the framework we lay out here is general and applicable to any relevant biological modeling setting where experimental data is obtainable. 

Formally, the contributions of this work are threefold:  
(1) Applying Bayesian optimal experimental design and high-performance computing (HPC) to a large-scale coupled ODE system of 23 equations and with 11 uncertain parameters and choosing between 5 experimental designs --- the largest PD system considered thus far in a fully Bayesian OED framework of which the authors are aware; (2) Proposing novel decision-relevant metrics for quantifying the uncertainty in therapeutic performance of novel inhibitor drugs at a given concentration (\ic); and (3) Identifying optimal experiments that minimize the uncertainty in therapeutic performance (i.e., simulated lethality) as a function of the inhibitor dosage of interest. 

As we will show, there is preference for measuring activated caspases at low \ic to reduce uncertainty in the probability of cell death. At larger \ic, our results show that this uncertainty is maximally reduced by collecting data the mRNA-Bax concentrations. Therefore, we conclude that the decision as to which species to experimentally measure must be selected by considering trade-offs between the predicted BOED objectives and the inhibitor viability \textit{in vitro}. The approach and results presented here thus bridge model-guided optimal design with uncertainty quantification, advancing drug discovery through uncertainty-aware decision-making.



\section{Methods}\label{sec:methods}

The problem addressed in this work is to understand the impact of model parameter uncertainty on the predictive capacity of the PD model for \textit{PARP1}-inhibited cellular apoptosis; a representative model for cancer drug evaluation. 
And given the effects of uncertainty and lack of experimental data, we wish reestablish a predictive model by recommending measurements that can be made in the lab to better constrain the model parameters. 

The workflow to achieve the stated goals is as follows: 

\begin{enumerate}
    \item Acquire many (synthetic) experimental measurements for each prospective measurable species in the model
    \item Construct \textit{prior probability distributions} (i.e., beliefs regarding model parameter distributions in the absence of data)
    \item Conduct parameter estimation via Bayesian inference using the model, data, and prior probabilities for each prospective experiment and over multiple data samples
    \item Compute expected drug performance predicted by the model given \textit{posterior probability distributions} (i.e., updated parameter distribution beliefs after incorporating data) 
    \item Rank and recommend experiments based on a metric that quantifies reliability in model predictions.
\end{enumerate}

\begin{figure}[H]
\centering
\begin{subfigure}[b]{0.85\textwidth}
    \adjustbox{trim=0pt 30pt 0pt 30pt,clip}{\includegraphics[width=\linewidth]{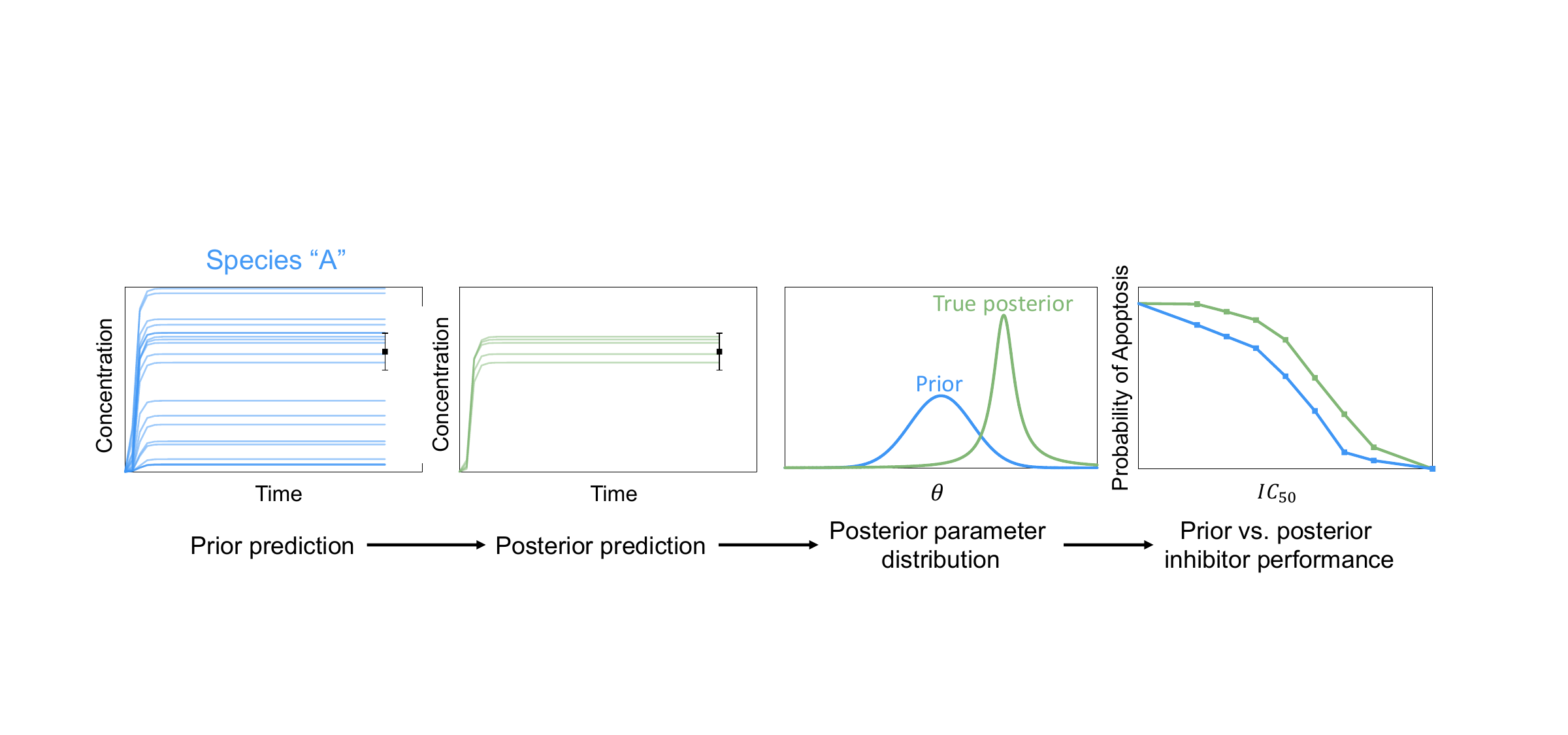}}
    \subcaption{Model parameter calibration with real data.}\label{fig:calibration_real}
\end{subfigure}
\vfill
\centering
\begin{subfigure}[b]{0.85\textwidth}
    \adjustbox{trim=0pt 10pt 0pt 10pt,clip}{\includegraphics[width=\linewidth]{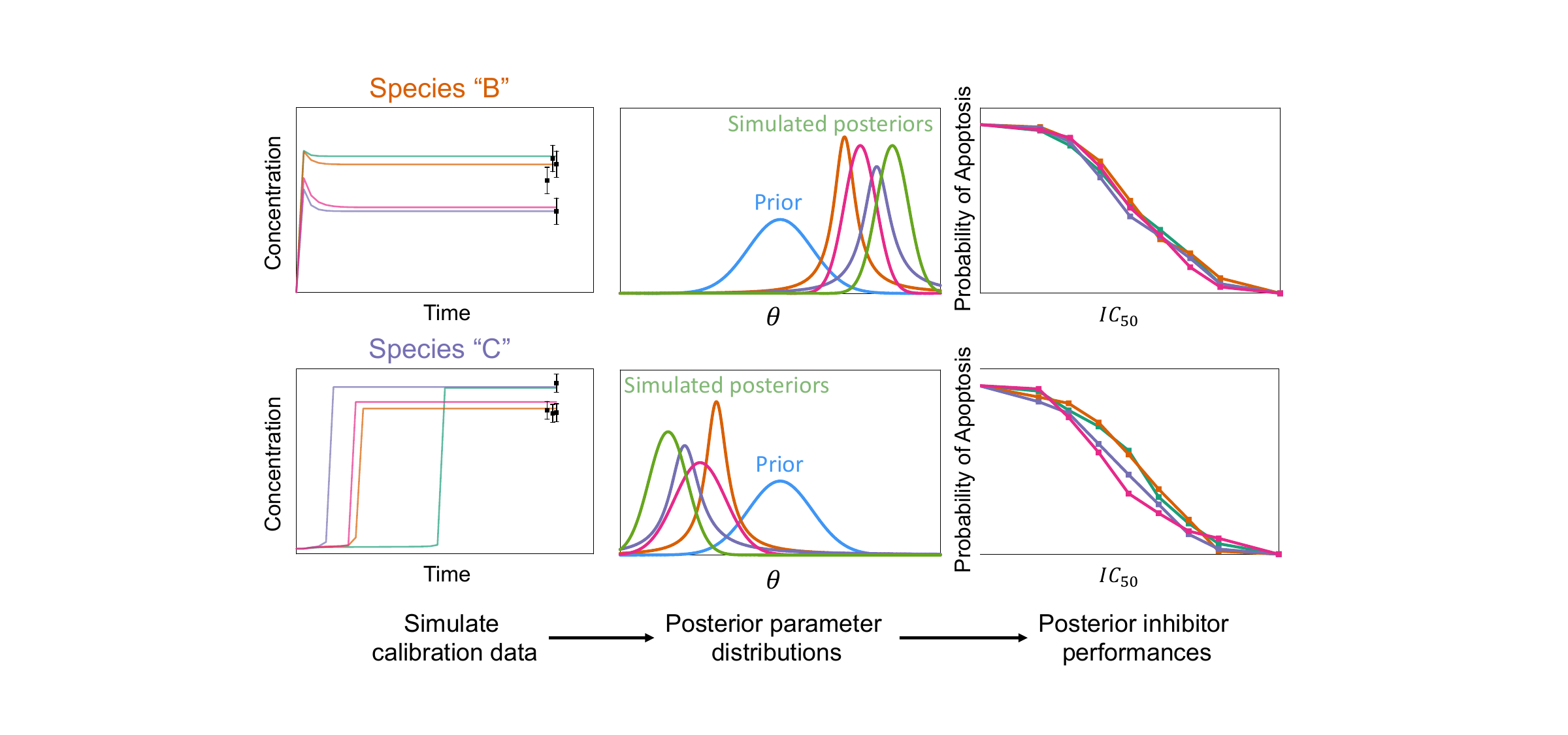}}
    \subcaption{Model parameter calibration with simulated experimental data.}\label{fig:calibration_sim}
\end{subfigure}
\caption{Workflow for Bayesian parameter calibration using (a) real, measured data and (b) experimental data simulated from the model.}
\label{fig:workflow}

\end{figure}

A graphical depiction comparing traditional parameter calibration to the proposed approach using simulated experimental data is shown in Figure~\ref{fig:workflow}. 

Traditional model parameter calibration via Bayesian inference is shown in Figure~\ref{fig:calibration_real}. 
The model-derived prior predictive distribution of concentration profiles for \textcolor{blue1}{Species ``A''} is constrained by experimental data (in \textbf{black}). 
This leads to an updated probability density function for the model parameters $\theta$ referred to as the \textit{posterior distribution}. 
The posterior distribution is then used to quantify inhibitor performance via probability of achieving apoptosis in simulation across an array of \ic values.
We can then compare the expected inhibitor drug performance between the prior and posterior uncertainty.
More accurate predictions by the posterior are thus expected because real data was used to update the model. 

The methodology utilized in the present work is presented in Figure~\ref{fig:calibration_sim}. 
Here, we do not have access to real, measured data to constrain our model. 
However, we can use the PD model and an expert-derived error model to generate an ensemble of simulated experimental data. 
In simulating and calibrating to a large volume of simulated experimental data, we account for uncertainty in what \textit{could} be measured in reality (in \textbf{black}). 
This then leads to an ensemble of posterior distribution predictions for the parameters $\theta$.
We obtain ensembles of posterior parameter distributions for several potential measurable species in the model (e.g., \textcolor{orange1}{Species ``B''} and \textcolor{purple1}{Species ``C''}). 
From these, we can compute an ensemble of inhibitor performances as measured by probability of apoptosis across a range of \ic values. 
This implies a distribution on expected inhibitor performances as predicted by ``measuring'' different species and using the simulated measurements for calibration. 
By comparing the distributions in expected inhibitor performance across different measured species, we can identify which experiment is ``optimal'' by selecting the one that leads to the greatest reduction in uncertainty.

\subsection{Mathematical Model}

The general form of a pharmakodynamic model is shown in Eqn.~\ref{eqn:math_model}, which represents a set of coupled ordinary differential equations. 

\begin{equation}\label{eqn:math_model}
\begin{alignedat}{2}
    \displaystyle & \dot{y}(t) = f\left(y(t); u, \kappa\right)\\
    & y(t_0) = y_0 &
\end{alignedat}
\end{equation}

In the rate-based modeling framework used here, variables $y(t) \in \mathbb{R}^n$ are state variables representing the time-varying concentrations of the $n$ chemical species of interest within the cell signalling pathway.
The parameters $\kappa \in \mathbb{R}^m$ are the parameter data, which includes the kinetic rate constants for the reaction network. 
Initial concentrations of species are those at $t_0$, specified by the value of the vector ${y}_0 \in \mathbb{R}^n$.
The system parameters and time-varying species concentrations are related by the functions $f : \mathbb{R}^n \rightarrow \mathbb{R}$, indexed by $\forall i \in \{1,\dots,n\}$. Also, we separate the input parameters in Equation~\ref{eqn:math_model} into two vectors: $u$, the control variables, and $\kappa$ the uncertain model parameters. The pre-specified input parameters $u$ are any experimental controls set \textit{a priori}. In the model considered here, this is the half maximal inhibitory concentration (\ic) for a novel \textit{PARP1} inhibitor molecule. \ic is a commonly used metric in experimental biology for quantifying a molecule's ability to inhibit a specific biological function.

\subsection{Bayesian Optimal Experimental Design}

Bayesian optimal experimental design is a statistical framework that aims to identify the most informative experimental conditions or measurements to improve parameter estimation and model predictions in the presence of uncertainty \cite{overstall2019bayesian, rainforth2023modern}. 
It involves iteratively selecting experiments \cite{dehghannasiri2014optimal,hong2021optimal} that maximize the expected information gain or other objectives, such as mean objective cost of uncertainty \cite{yoon2013mocu, yoon2021mocu}, given prior knowledge, model uncertainty, and the available resources. 
The goal is to strategically design experiments that provide the most valuable information via uncertainty reduction in the calibrated model predictions. 

In our application, we want to identify an experimental design $\xi$ (i.e., a specific species we can measure from the cell in the presence of a \textit{PARP1} inhibitor) from the set of feasible designs, $\Xi$, which optimizes information gain regarding uncertain parameters, $\theta$, given the outcome of that experiment, $y$. 
Let us denote the set of all uncertain parameters within the ODE system, including initial conditions and rate constants, as the set $\Theta$. 
For tractability reasons, we may choose to only consider a subset of the parameters which are deemed most significant via global sensitivity analysis and refer to the vector of key, significant parameters as $\theta \in \Theta$. 
We also denote the vector of experimental designs being considered as $\xi \in \mathcal{X} \subset \Xi$, and $|\mathcal{X}| = N_{\xi}$ is the number of design decisions to choose from in optimization. 
Here, the individual experiments represent a species to measure in an experiment representing the process modeled in the \textit{PARP1}-inhibited cell apoptosis PD model. 
We assume prior distributions on the uncertain parameters $p(\theta)$, and a likelihood of the data conditional on the parameters and controls variables, $p(y|\theta, \xi, u)$. 
For a general objective function of interest, $\mathcal{J}(\xi; u)$, where $\xi$ are the decision variables and $u$ inputs to Equation~\ref{eqn:math_model}, the Bayesian optimal experimental design task is described by Equation~\ref{eqn:boed}.

\begin{alignat}{1}
    \xi^{\ast} = \arg \max_{\xi \in \Xi} \mathcal{J}(\xi;u)  \label{eqn:boed}
\end{alignat}

Traditionally, this objective $\mathcal{J}$ is the expected information gain (EIG), a quantitative measure of entropy reduction between the prior and posterior probability distributions of $\theta$\cite{lindley1956measure, huan2013simulation}. 
Under such a formulation, the objective function becomes a nested expectation computation under the marginal distribution of experimental outcomes. 
The posterior distributions are often then estimated using a Markov Chain Monte Carlo (MCMC) algorithm to compute the information gained.
For this application, we are not interested in optimal experiments which simply minimize relative entropy in the prior and posterior distributions in $\theta$, as is the standard practice in BOED.
Instead, we wish to select optimal experimental designs that minimize uncertainty in lethality predictions made by the ODE model under different inhibitor drug \ic, i.e., different $u \in \mathcal{U}$. 
In this model setting, lethality (i.e., probability of cell apoptosis) can be viewed as a probabilistic analog to the traditional dose-response curve and is defined in Equation~\ref{eqn:p_apop}. 

\begin{figure}
\centering
\begin{subfigure}[b]{0.5\textwidth}
\centering
    \includegraphics[width=\linewidth]{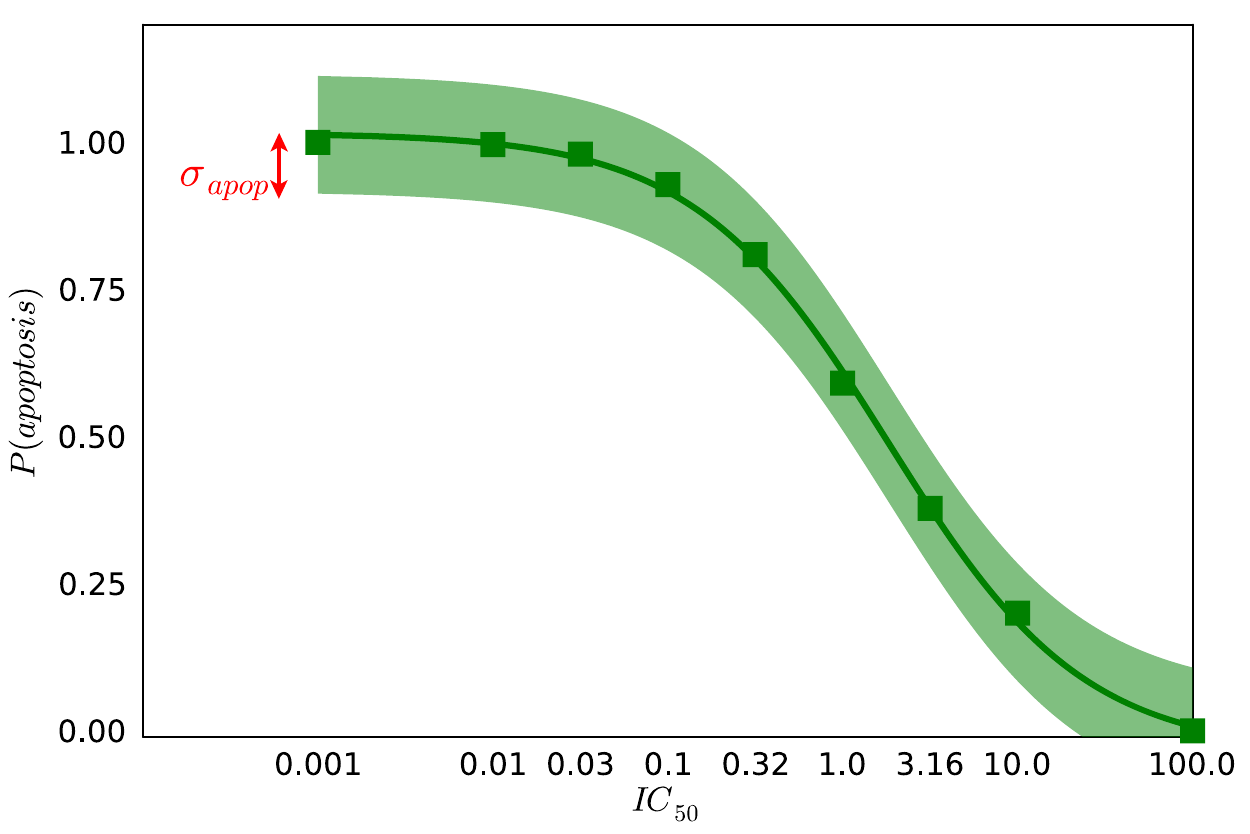}
    \caption{}
    \label{fig:sigma_apop}
\end{subfigure}
\vfill
\begin{subfigure}[b]{0.5\textwidth}
\centering
    \includegraphics[width=\linewidth]{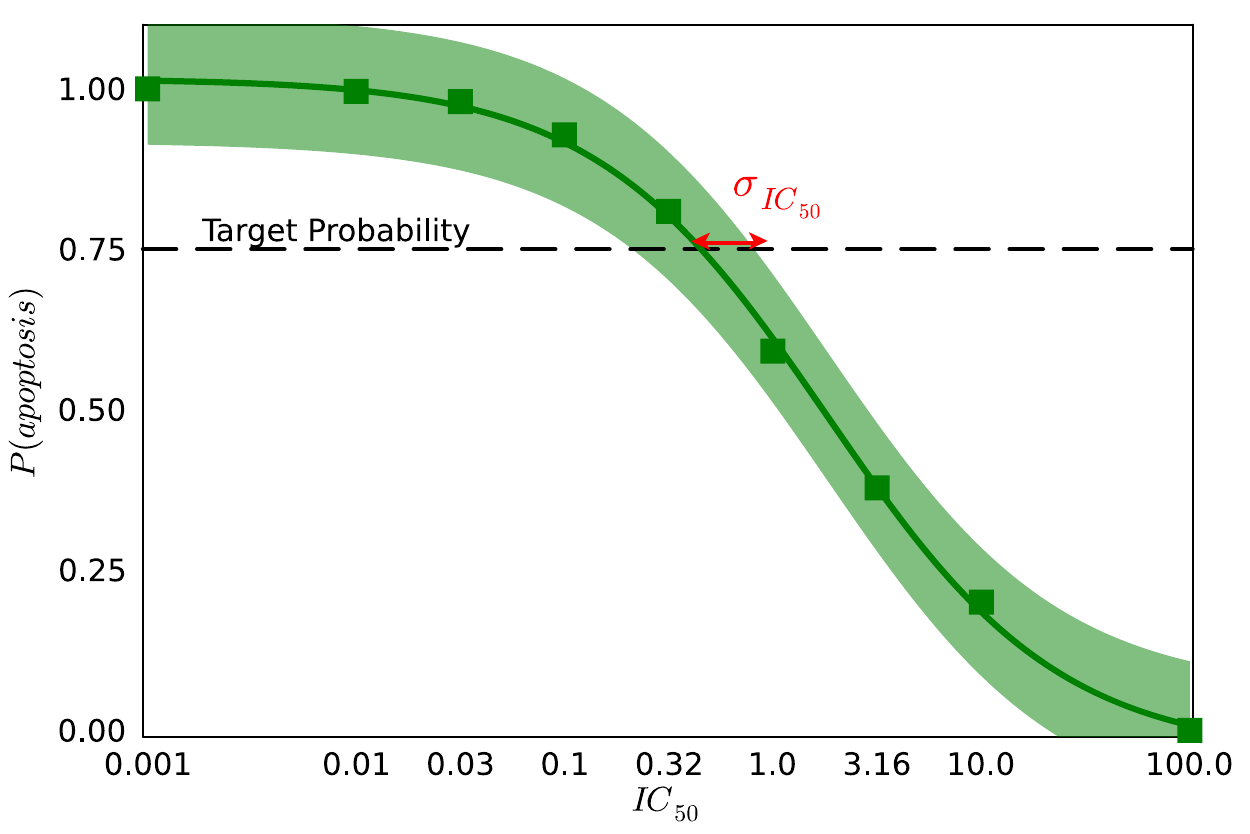}
    \caption{}
    \label{fig:sigma_ic50}
\end{subfigure}
\caption{Graphical depictions of uncertainty metrics for probabilistic lethality curves in (a) $\sigma_{apop}(u)$ at $u=0.001$ and (b) $\sigma_{IC_{50}}(\mathbb{T})$ at target probability threshold $\mathbb{T}$ = 0.75. The shaded green region represents a distribution of \textit{lethality curves} that is computed from an ensemble of posterior distributions in the uncertain parameters $\theta$. The mean of this distribution in lethality curves is shown in the square data points.}
\label{fig:lethality_metrics}
\end{figure}

To that end, we devise two metrics for quantifying uncertainty in cell lethality predictions at different \ic values. 
The first is the uncertainty in the probability of triggering cell apoptosis at a given \ic, represented as $\sigma_{apop}(u)$.  
A graphical representation of $\sigma_{apop}(u)$ across \ic is shown in Figure~\ref{fig:sigma_apop}.
This quantity defines a characteristic height of the distribution of the lethality curve predictions at a specific \ic of interest.
Smaller $\sigma_{apop}$ values thus imply that there is higher confidence in the PD model predictions at the \ic they are computed at. 
One may choose to only consider the $\sigma_{apop}$ at very small \ic, since we are interested in drug molecules which are most effective at low doses. 

The second metric provides a quantification of the uncertainty in the inhibitor drug dosage (i.e., \ic) that achieves a given probability of triggering cell apoptosis, and is represented as $\sigma_{IC_{50}}(\mathbb{T})$. 
A graphical representation of of $\sigma_{IC_{50}}(\mathbb{T})$ across \ic and for $\mathbb{T} = 0.75$ is shown in Figure~\ref{fig:sigma_ic50}.
This metric defines a characteristic width of the lethality distribution at a target probability of triggering cell death, $\mathbb{T} \in [0,1]$. 
A smaller $\sigma_{IC_{50}}$ value at a given $\mathbb{T}$ implies higher confidence in which \ic value achieves a given probability of killing the cell.
More detail on these metrics is discussed in Appendix~\ref{STAR:voi_metrics}.
In addition, we discuss the multi-objective treatment of these two metrics, and the sensitivity of the optimal recommended experiment to that treatment, in Section~\ref{sec:results}.

\section{Application}\label{sec:application}

We consider a dynamic \textit{PARP1}-inhibited cell apoptosis models for both wild-type human cells published in \cite{Mertins2023.02.08.527527}.
We note that this pathway can be found in some tumorous cells, as well.
Thus, for the present study, the process that this model represents is the inhibition of \textit{PARP1} as a therapy for cancer.
By applying an inhibitor via a specified \ic, inhibition of \textit{PARP1} halts the DNA strand break repair process, reducing damage repair pathways and potentially leading to cell apoptosis. 
Further information on the model and the complete system of differential equations is available in the Equation S\ref{appendix:ode_model_full}.

The model consists of a set of 23 ordinary differential equations (ODE) wherein a subset of the 27 rate constant parameters are considered uncertain. 
This model takes as inputs (1) an \ic value and (2) initial conditions and (3) kinetic parameter values. 
In the present study, we only consider output at a specific time point, (i.e., the final time point) when performing inference. 
This means that the likelihood we supply for the data follows a Normal distribution in the model prediction at $t=\tau$, where $\tau$ is the time limit for the numerical integrator used to solve the ODE system at a given \ic, and with a known $\sigma$ that is computed with a $10\%$ error relative to the model prediction. 
Information for the uncertain parameters considered in this study was first elicited from expert understanding of the system and relevant sources.
Nominal parameter values and confidence interval information is shown in Table S\ref{table:priors}. 
Details regarding the construction of all prior distributions on uncertain parameters considered in the present study are provided in Table S\ref{table:prior_distributions}. 
In addition, because experimental data for species concentrations within a single cell undergoing \textit{PARP1}-inhibited apoptosis are not readily available, we utilize simulated experimental data for our study that is generated via the procedure described in Appendix Section~\ref{STAR:data}. 

\subsection{Objective function specification}\label{sec:objective}
As noted previously, the quantity-of-interest in assessing the performance of \textit{PARP1} inhibitor is the predicted lethality.
The predicted lethalities are then used to compute our uncertainty metrics or objective functions, $\sigma_{apop}(u)$ and $\sigma_{IC_{50}}(\mathbb{T})$.
Here, we specify the method for computing lethality, i.e., probability of triggering cell death.
It has been shown that a activated caspase molecule count in the cell above a threshold of 1,000 is sufficient to trigger cell apoptosis \cite{liepe2013maximizing}.
Therefore, we can compute whether or not cell apoptosis has been irreversibly triggered by the introduction of a \textit{PARP1} inhibitor by computing $y_{k}^{max} := \displaystyle \max_{t \in [1,\dots,\tau]} y_{k}(t)$ where $k=$Casp-act.
Thus, lethality, or $P(apoptosis)$ can be computed by sampling the prior (posterior) probability distributions of the uncertain parameters, $\theta^i \sim p(\theta) \ \forall i \in [1,\dots, N_s]$, specifying a value for \ic, and integrating the system of ODEs. Given the ensemble of $N_s$ maximum-caspase concentrations, $P(apoptosis)$ can be calculated as shown in Equation~\ref{eqn:p_apop}.

\begin{alignat}{2}
    & P(apoptosis) := \frac{1}{N_{s}} \sum_{i=1}^{N_s} \mathbbm{1}_{y_{Casp,i}^{max} \geq 1000} \label{eqn:p_apop}
\end{alignat}

Here, $N_s$ is the Monte Carlo sample count.
When computed at all \ic of interest, we retrieve a vector of $P(apoptosis|u)$ predictions $\forall u \in \mathcal{U}$, producing a \textit{lethality curve} for a given posterior parameter estimate.




\subsection{Incorporating biological constraints}\label{sec:constraints}

In addition to the expert-elicited prior probability information in Table S\ref{table:priors}, there is other biologically and therapeutically relevant information that we can incorporate into Bayesian parameter estimation to further constrain our posterior predictions. 
First, we know that for \textit{PARP1} inhibitor drugs with sufficiently high \ic values (i.e., low efficacy), we expect the repair signal from \textit{PARP1} to win out over the programmed cell death induced by p53. 
This means that at high \ic, the DNA double strand break will be repaired and the cell will live. 
We also know the opposite to be true: for low \ic, programmed cell death will be induced and the cell will die. In mathematical terms, we can say:

\begin{align}\label{eqn:constraints}
    P(\textrm{apoptosis}) =
    \begin{cases} 0,&\text{if \ic} \geq 100\\
     1, &\text{if \ic} \leq 0.001\\
        \textrm{Equation}~\ref{eqn:p_apop} & \textrm{otherwise}
    \end{cases}
\end{align}

This prior knowledge regarding the effect of a drug on our quantity of interest does not directly translate to knowledge regarding the uncertain model parameters. 
However, because the quantity of interest is computed directly from an output of the model (i.e., concentration of activated caspases) we can still include this prior knowledge in Bayesian inference in three equivalent ways: in the likelihood, the prior, or through \textit{post facto} filtering of the estimated posterior chain. 
For the present application, we choose the posterior filtering approach due to numerical challenges in the HMC sampling algorithm used to estimate the posterior distribution. 

The posterior filtering approach used involves sampling a larger chain (in this study, 50,000 samples) for each simulated experimental data point.
If a given parameter sample in the posterior chain leads to the violation of the constraints in Equation~\ref{eqn:constraints} when specified in the forward simulation model under the nominal \ic (i.e., Talazoparib, see Appendix Section~\ref{STAR:data}), that sample is removed from the chain.


\section{Results and Discussion}\label{sec:results}

In this section, we provide the key findings of our study applying Bayesian optimal experimental design to the \textit{PARP1}-inhibited cell apoptosis model. First, we conduct a prior sensitivity analysis, follow by posterior estimation and computation of our optimal experimental design objectives. 

\subsection{Prior Sensitivity}

\begin{figure}
\centering
\begin{subfigure}[b]{0.45\textwidth}
    \includegraphics[width=\textwidth]{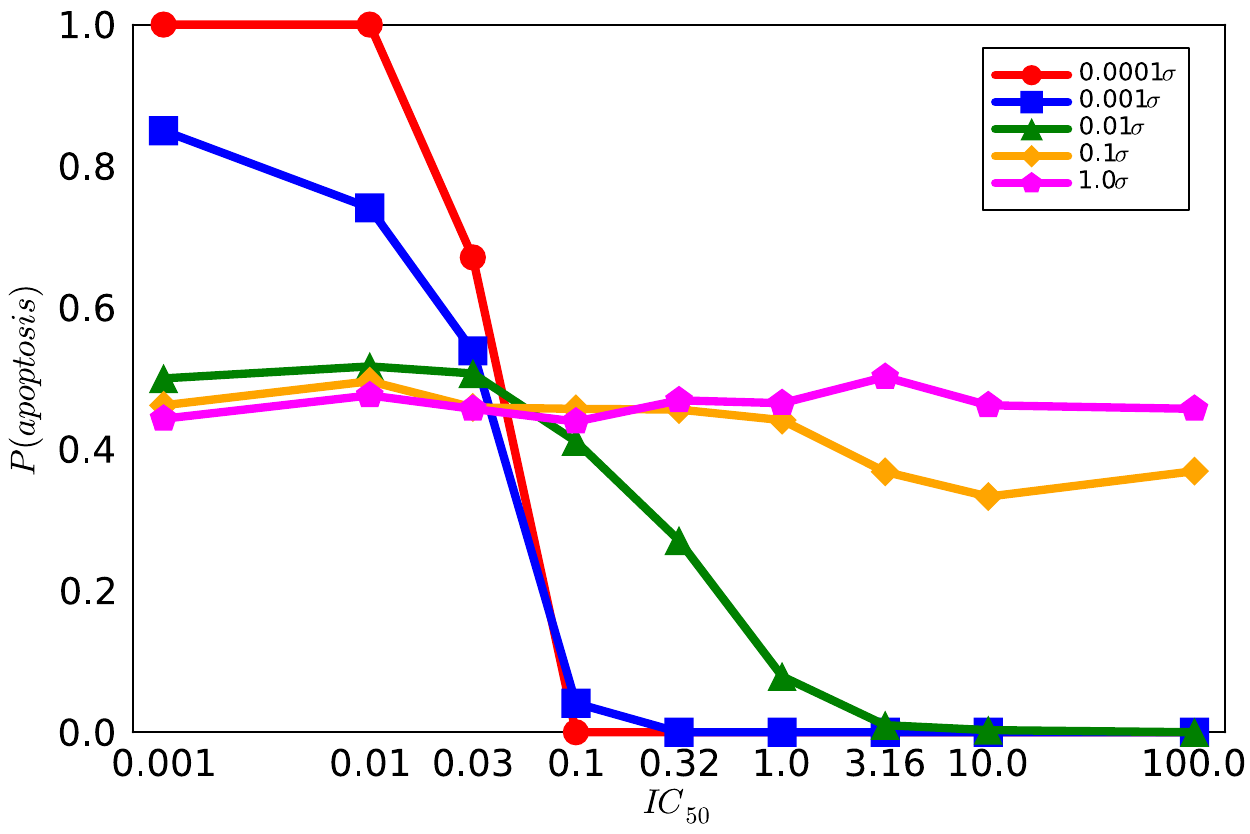}
    \subcaption{}\label{fig:unconstr_prior_SA}
\end{subfigure}
\hfill
\begin{subfigure}[b]{0.45\textwidth}
    \includegraphics[width=\textwidth]{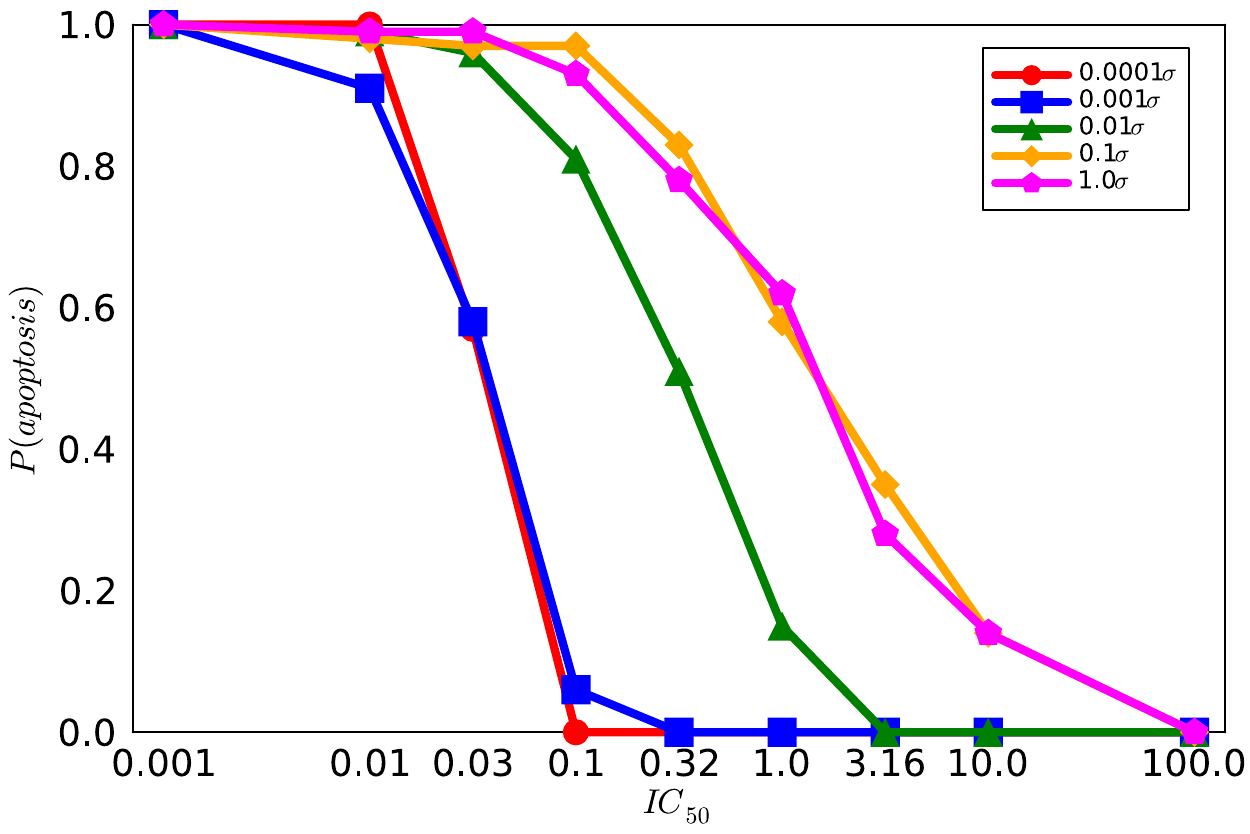}
    \subcaption{}\label{fig:constr_prior_SA}
\end{subfigure}
\caption{The probability of cell death $P(\text{apoptosis})$ plotted against \ic for different multiplicative factors of the standard deviation $\sigma$ in the prior distributions (a) without and (b) with applying the constraints in Equation~\ref{eqn:constraints}.}
\label{fig:prior_sensitivity}
\end{figure}

We conducted a prior sensitivity analysis computed for a quantity of interest, $P(\text{apoptosis})$. 
The purpose of this study is to understand how much reduction in uncertainty is required to achieve desired lethality curve predictions. The result of this study is shown in Figure~\ref{fig:prior_sensitivity}. 
Here, the $P(\text{apoptosis})$ is computed by (1) drawing 1,000 random samples from prior probability distributions for each uncertain parameter, then (2) solving the model in Equation~S\ref{appendix:ode_model_full} at those parameter sample values and (3) computing $P(\text{apoptosis})$ via Equation~\ref{eqn:p_apop} for all \ic = [0.001, 0.01, 0.032, 0.1, 0.32, 1.0, 3.2, 10.0, 100.0]. 
To determine the model sensitivity against the priors, we scale $\sigma$ for each parameter by a multiplicative factor to narrow the support of the prior distribution around its nominal mean. 
In Figure~\ref{fig:unconstr_prior_SA}, we neglect our prior constraints and just solve the model under the unconstrained prior samples. 
Under these conditions, we see that given the nominal prior uncertainty (i.e., $1.0\sigma)$, the effect of \ic on lethality vanishes (i.e., it is a more-or-less flat line). From the results, it appears that the model requires \textit{at minimum} a reduction in prior uncertainty of two orders of magnitude (i.e., $0.01\sigma$) before a predictive model is restored via sigmoidal curvature in the prior lethality curve. 
This implies that achieving a reduction in uncertainty spanning two orders of magnitude across all uncertain parameters is imperative to differentiate therapeutic performance between different inhibitors using the PD model.

In Figure~\ref{fig:constr_prior_SA}, we conduct the same sensitivity analysis while also applying the constraints in Equation~\ref{eqn:constraints} to restrict the model-predicted lethality values using biological insights. 
Under these conditions, we obtain lethality curve predictions that have the desired sigmoidal shape, showing low therapeutic performance at large \ic and high therapeutic performance at small \ic. 
However, it is also evident that significant reduction in parameter uncertainty (0.0001$\sigma$) is required to achieve apoptotic switch behavior at low \ic.\cite{liepe2013maximizing}   

It is worth noting that this sensitivity analysis involved altering all prior $\sigma$ values by the same amount, and the existence of parameter correlations could potentially yield alternative outcomes in terms of the necessary uncertainty reduction. However. these results underscores the need for substantial reduction in parameter uncertainty in PD models to re-establish predictive ability.

\subsection{Posterior Estimation}

We estimate posterior probability distributions for the uncertain parameters using the proposed Bayesian optimal experimental design approach. 
Details regarding the underlying algorithm are provided in Appendix Section~\ref{STAR:algorithm}. 
The summary statistics for (1) mean posterior parameter estimates, (2) effective sample size per parameter, and (3) filtered chain length are shown in Table S\ref{table:chain_data_mRNA_Bax}--S\ref{table:stat_casp_act}. 
The filtering procedure leads to non-uniform posterior chain lengths and lower effective sample sizes, as shown in the summary statistics. 
We note that only approximately 1\% of samples in the posterior chains pass the constraints enforced via Equation~\ref{eqn:constraints}. 
This is indicative of a known challenge in applying feasibility constraints after sampling: that the vast majority of the parameter space explored is ultimately infeasible, even for sufficiently large chains. 
Therefore, in future work, we aim to explore methods such as Bayesian constraint relaxation \cite{duan2020bayesian} to enforce relaxed constraints on our probabilistic model while maintaining numerical stability in standard Hamiltonian Monte Carlo sampling algorithms. 
Still, the average ESS per parameter is still close to the average chain length for each species, implying a high number of non-correlated samples in the posterior chains we computed. 
Additionally, we note that the assumption of Log-Normal prior probability distributions for all the model parameters in Table~S\ref{table:priors} is an assumption with ramifications on the final derived posterior distributions. 

\subsection{Value of Information Estimation}

\begin{figure}[H]
\centering
     \begin{subfigure}[b]{0.3\textwidth}
         \centering
         \includegraphics[width=\textwidth]{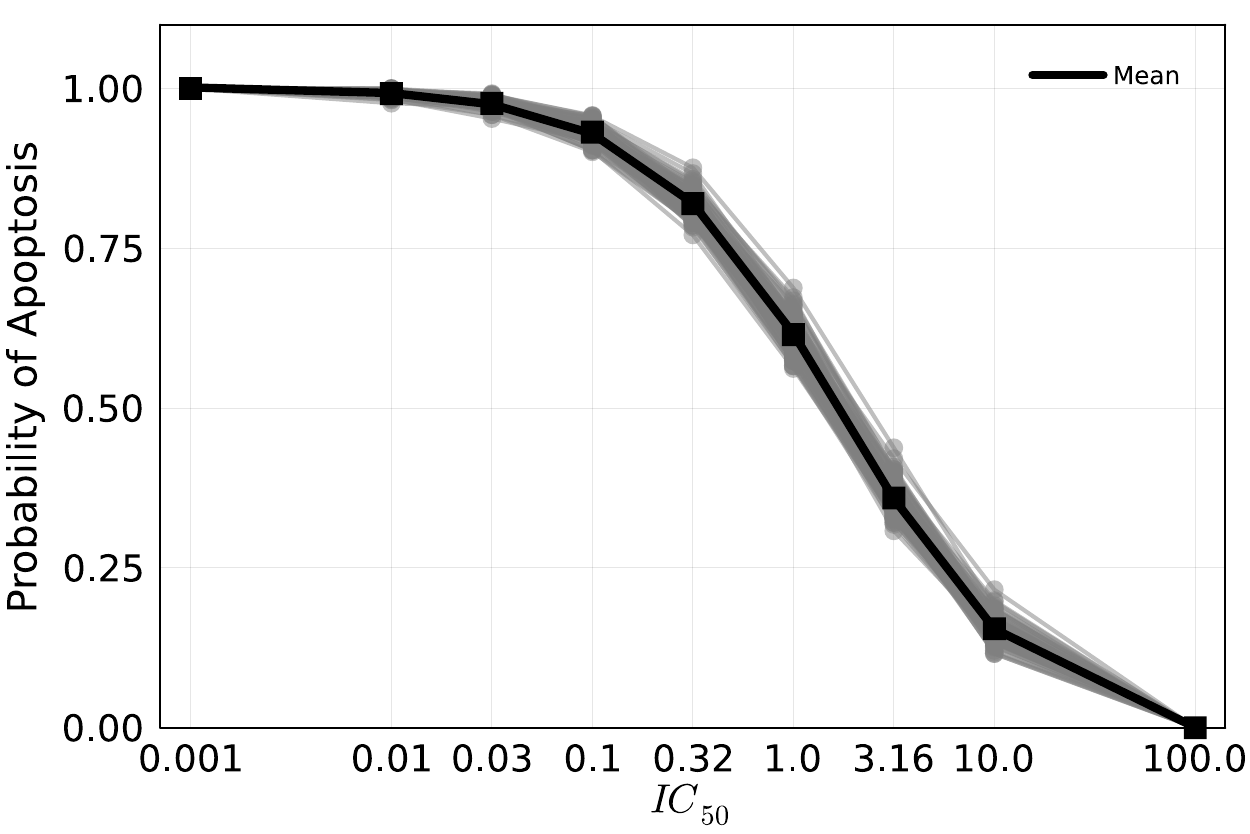}
         \caption{mRNA-Bax}
         \label{fig:lethality_13}
     \end{subfigure}
     \hfill
     \begin{subfigure}[b]{0.3\textwidth}
         \centering
         \includegraphics[width=\textwidth]{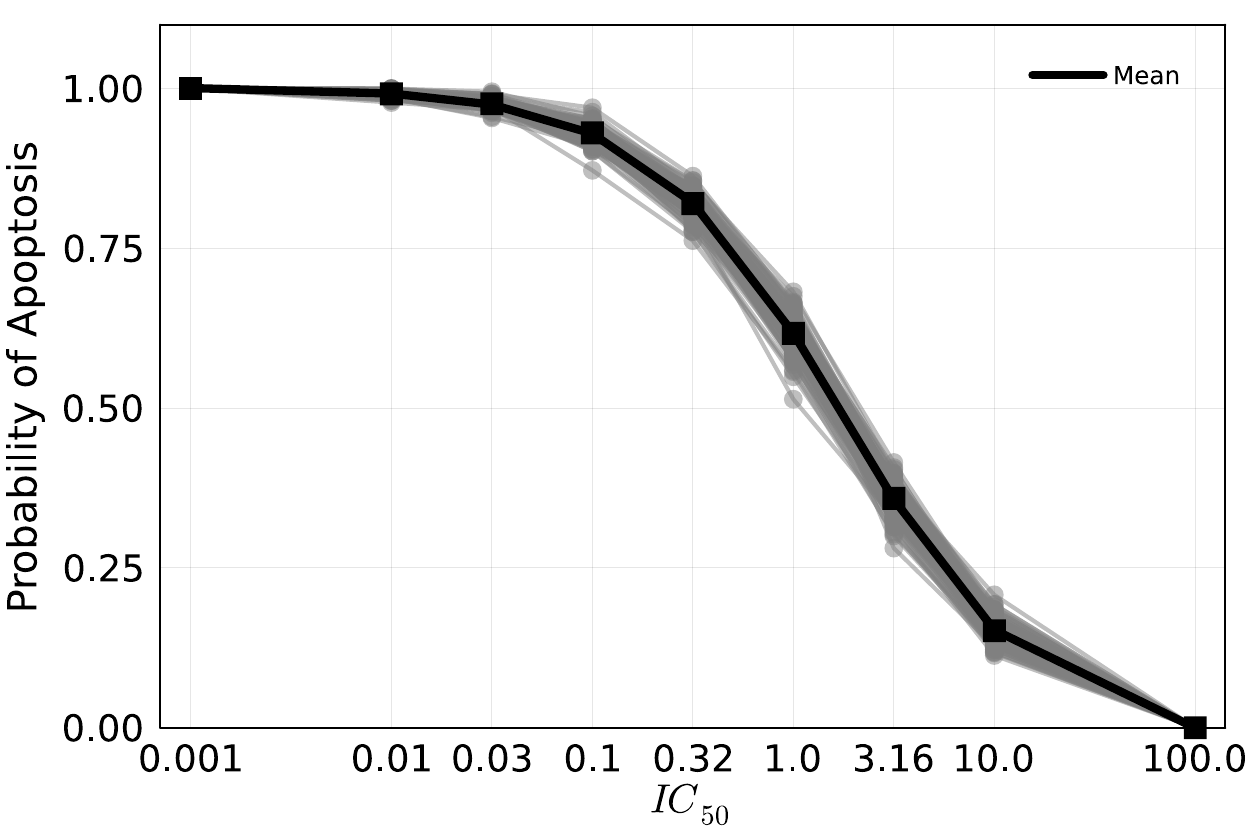}
        \caption{Bad-Bcl-xL}
        \label{fig:lethality_15}
     \end{subfigure}
     \hfill
     \begin{subfigure}[b]{0.3\textwidth}
         \centering
         \includegraphics[width=\textwidth]{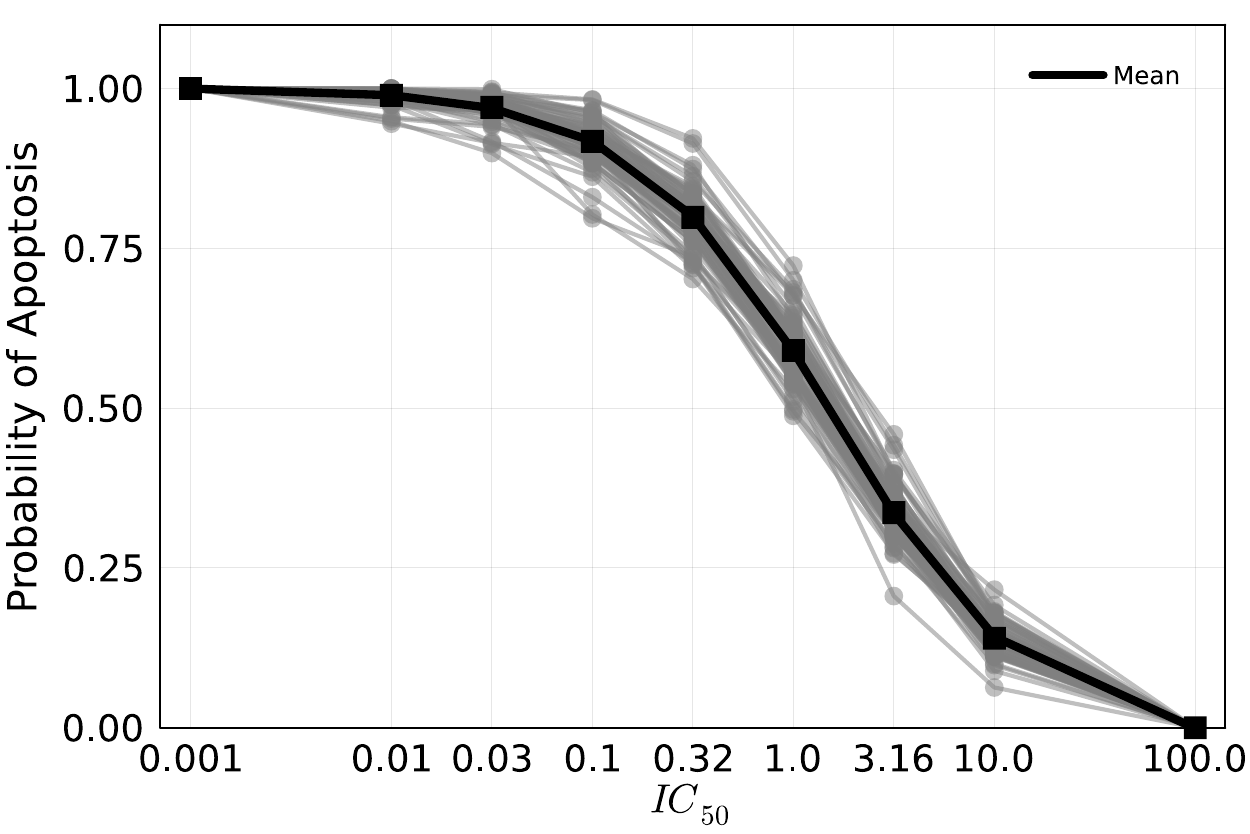}
         \caption{Casp-pro}
         \label{fig:lethality_17}
     \end{subfigure}
     \hfill
     \begin{subfigure}[b]{0.3\textwidth}
         \centering
         \includegraphics[width=\textwidth]{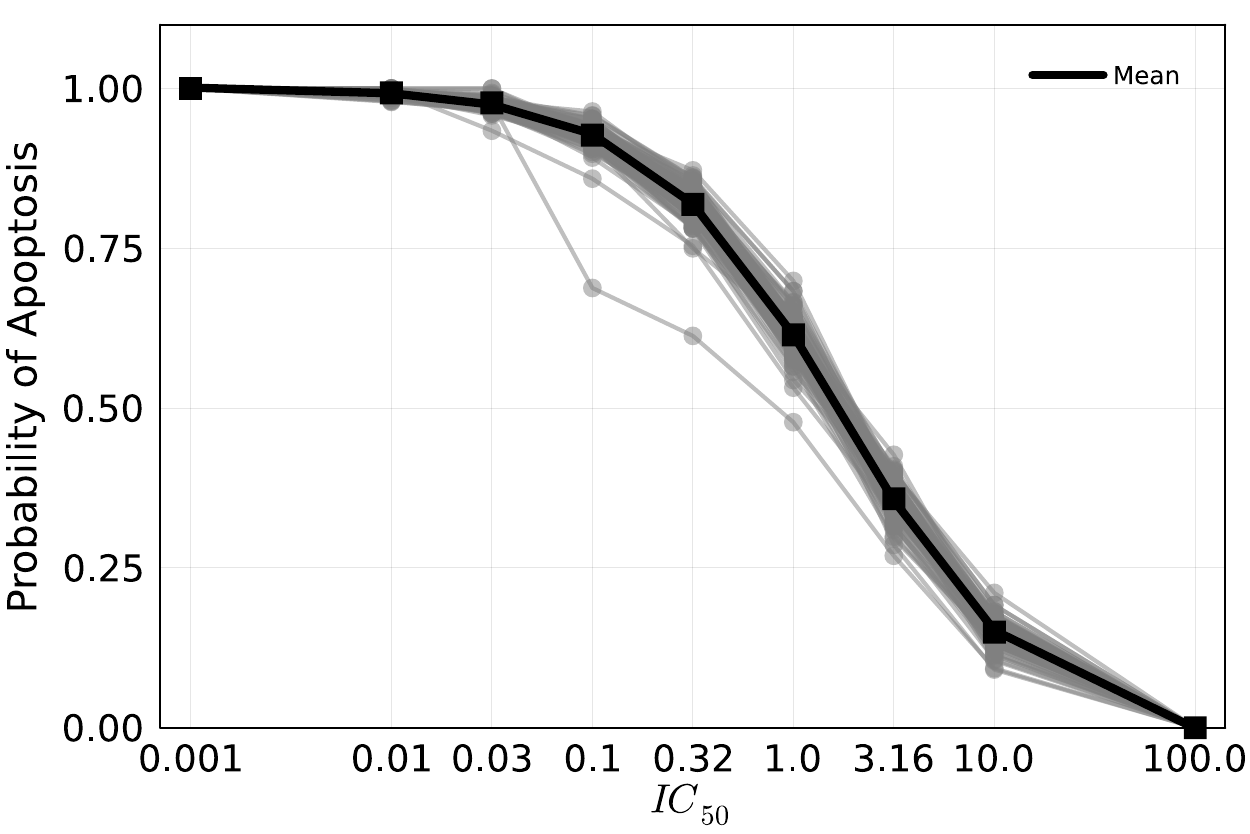}
         \caption{Bax-Bcl-xL}
         \label{fig:lethality_20}
     \end{subfigure}
     \hfill
     \begin{subfigure}[b]{0.3\textwidth}
         \centering
         \includegraphics[width=\textwidth]{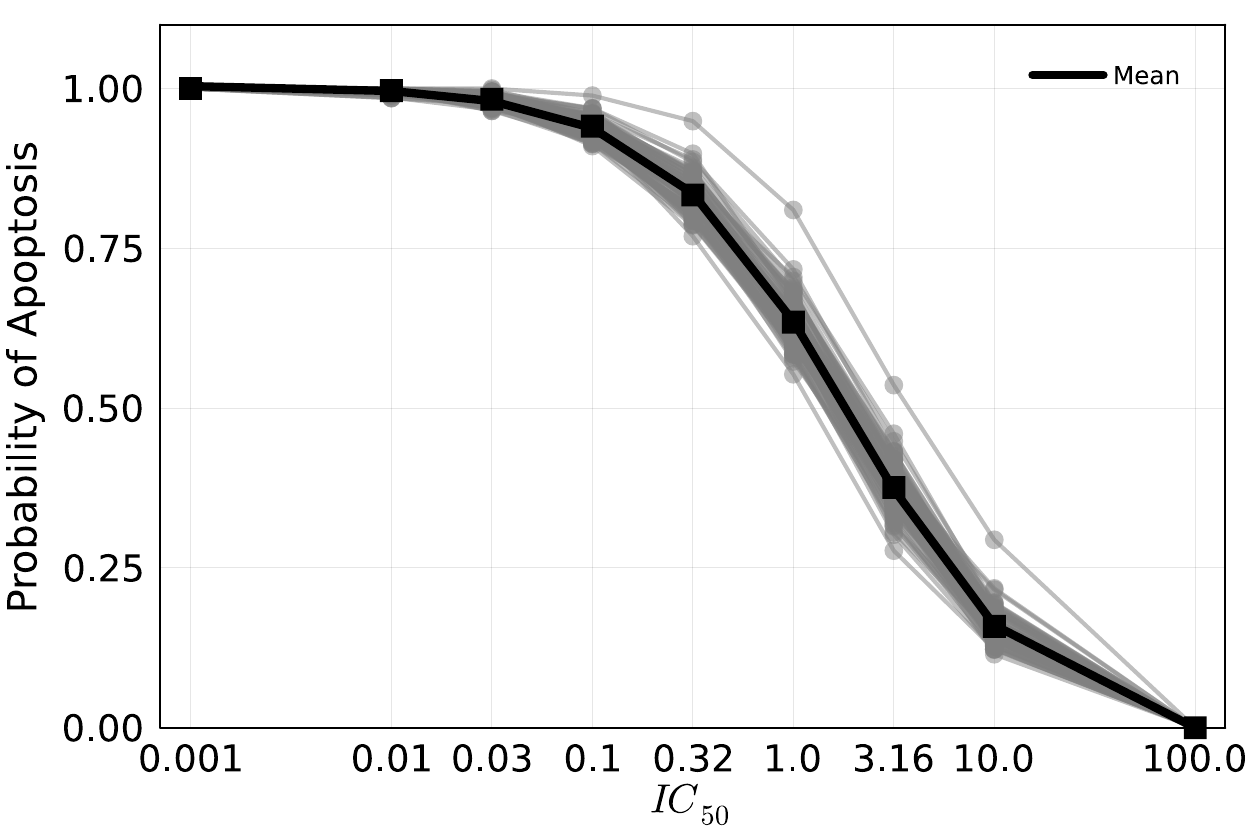}
         \caption{Casp-act}
         \label{fig:lethality_22}
     \end{subfigure}
        \caption{Ensemble of lethality curves predicted for each measurable species considered at different \ic values. Each individual curve represents a different lethality prediction obtained from a single synthetic experimental measurement. The mean over all 100 measurements is shown in black.}
        \label{fig:lethality_graphs}
\end{figure}

\begin{figure}[H]
\centering
    \begin{subfigure}[b]{0.4\textwidth}
         \centering
         \includegraphics[width=\textwidth]{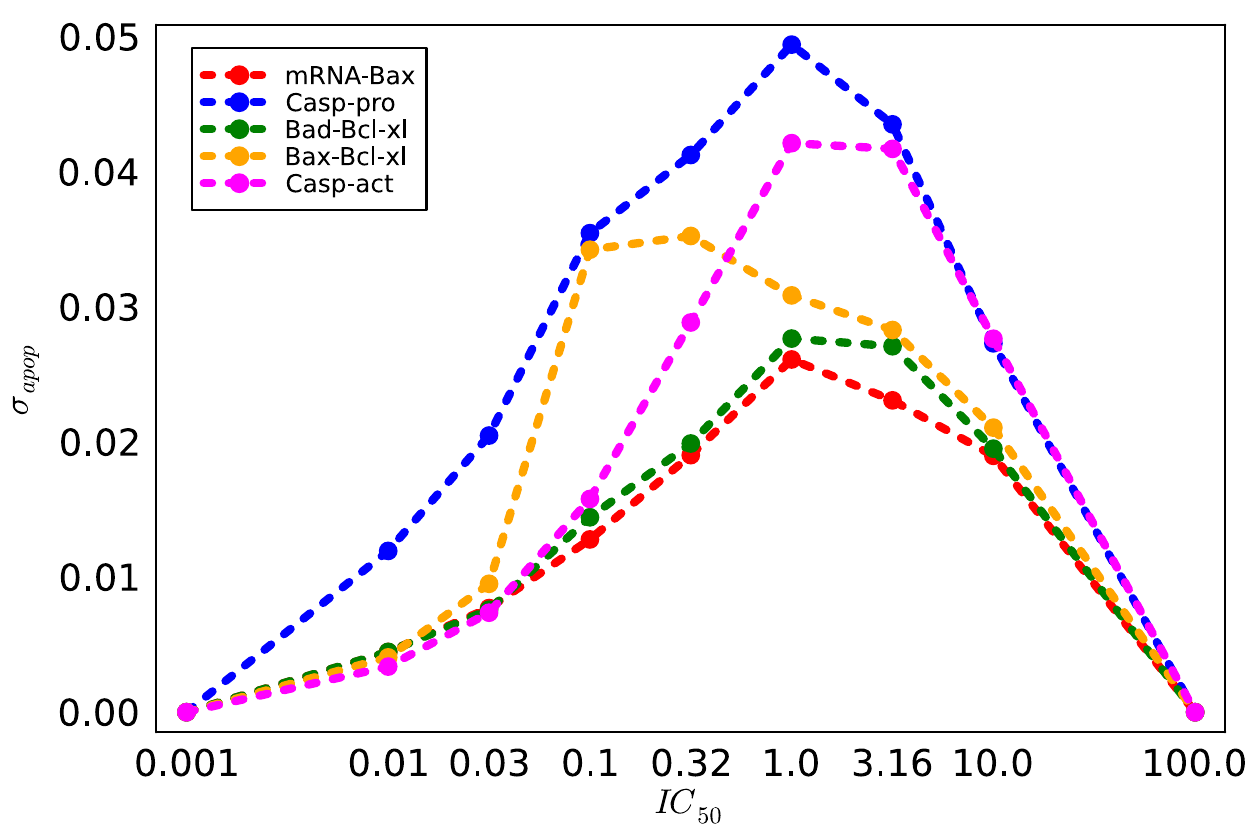}
        \caption{$\sigma_{apop}$}
        \label{fig:all_sigma_apop}
     \end{subfigure}
     \begin{subfigure}[b]{0.4\textwidth}
         \centering
         \includegraphics[width=\textwidth]{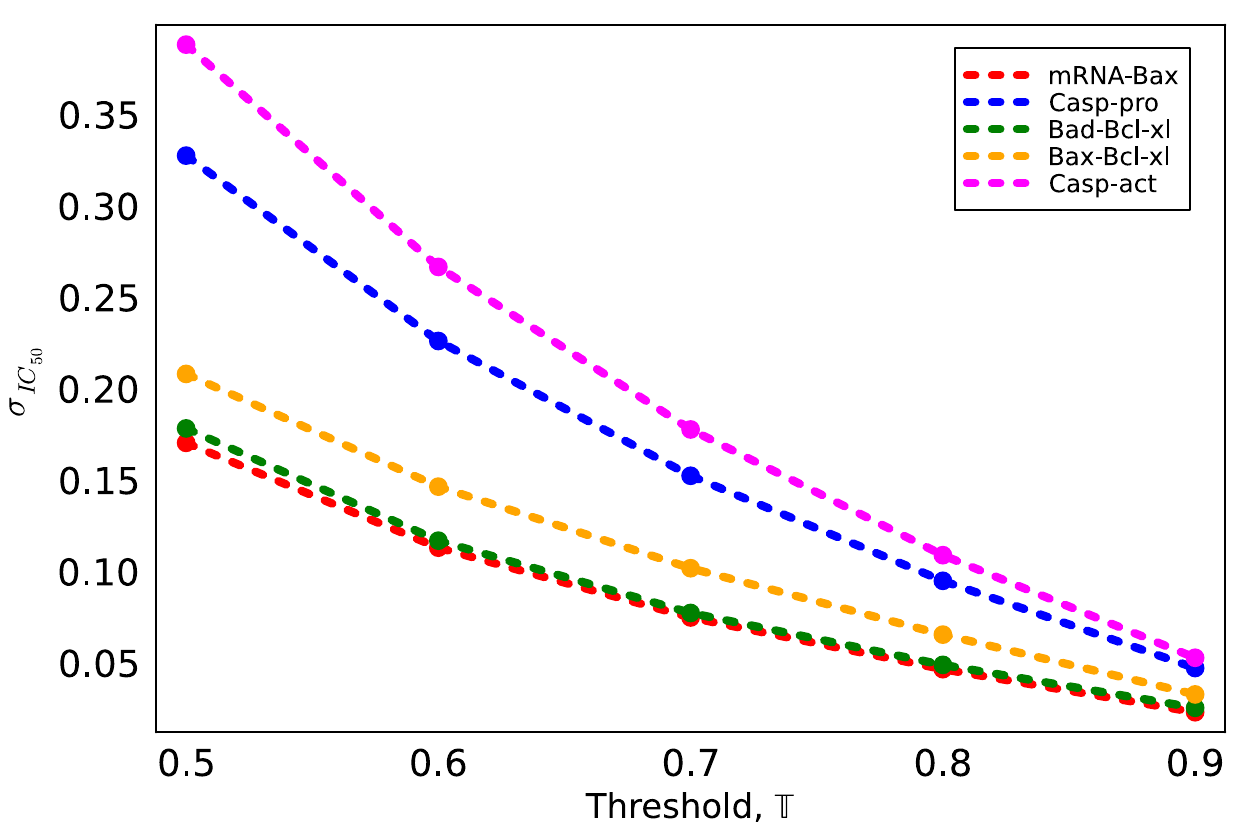}
         \caption{$\sigma_{IC_{50}}$}
         \label{fig:all_sigma_ic50}
     \end{subfigure}
     \hfill\caption{Uncertainty metrics for synthetic lethality computed at different \ic using posterior distributions inferred using concentration data for the measurable species considered. }
\end{figure}

The ensemble of predicted lethality curves for each species in Table~S\ref{table:observables} are shown in Figure~\ref{fig:lethality_graphs}.  
It is clear from the plots that certain species induce less spread in lethality predictions across \ic (e.g., mRNA-Bax) while other exhibit more variability (e.g., Casp-pro). 
To quantitatively understand these trends, we compute the the values of $\sigma_{apop}(u)$ and $\sigma_{IC_{50}}(\mathbb{T})$. The plots in Figures~\ref{fig:all_sigma_apop}-\ref{fig:all_sigma_ic50} show how these uncertainty metrics evolve over $u$ and $\mathbb{T}$, respectively.
Tabulated results are available in Tables~S\ref{table:sigma_apop} and S\ref{table:sigma_ic50}. 

It is seen in Figure~\ref{fig:all_sigma_apop} that there is a regime at small \ic  (which is the regime with the greatest therapeutic viability) wherein the minimizer of $\sigma_{apop}$ changes. 
In particular, at \ic = 0.01 and \ic = 0.03, the $\xi$ which minimizes the $\sigma_{apop}$ objective is Casp-act. 
At \ic > 0.03, the minimizer becomes mRNA-Bax for the rest of the \ic range considered. This can also be seen in Table~S\ref{table:sigma_apop}, where the minimum value of $\sigma_{apop}$ in each column (i.e., for each \ic) is demarcated with an $\ast$. 

When considering the $\sigma_{apop}$ objective alone, and when only placing significance in reducing uncertainty at low \ic, the optimal decision $\xi^{\ast}$ = Casp-act. 
At \ic = 0.01, the potential reduction in uncertainty in $P(apoptosis)$ under experimental designs (i.e., measurements) $\xi^{\ast} = $ Casp-act is 24\% when compared to Bad-Bcl-xL, the species with the highest $\sigma_{apop}(0.01)$. 
We note again that reduction in $\sigma_{apop}$ directly translates to a reduction in the uncertainty regarding the predicted $P(apoptosis)$ for a given inhibitor (i.e., at a given \ic).  
This is analogous to reducing uncertainty in a given inhibitor drug's expected performance. 
In summary, this result implies that it is optimal to acquire and calibrate to experimental data regarding Casp-act concentrations if we wish to maximize confidence in the expected lethalities of inhibitors with with low \ic. 

In the case of $\sigma_{IC_{50}}$, as shown in Figure~\ref{fig:all_sigma_ic50}, the trend of which $\xi$ minimizes the objective remains consistent across threshold values $\mathbb{T}=[0.5, 0.6, 0.7, 0.8, 0.9]$. 
Thus, when considering the $\sigma_{IC_{50}}$ objective alone, and for any threshold, the optimal $\xi^{\ast}$ = mRNA-Bax. 
We note that reduction in $\sigma_{IC_{50}}$ directly translates to a reduction in the uncertainty regarding which \ic achieves the target $P(apoptosis)$ threshold, $\mathbb{T}$. 
This is analogous to reducing uncertainty in a which inhibitor drug achieves the desired expected performance. 
For a potent inhibitor with $\mathbb{T} = $ 0.9, selecting the optimal experimental design $\xi^{\ast}$ = mRNA-Bax over Casp-act (i.e., the species with the highest $\sigma_{IC_{50}}(0.9)$) results in a substantial 57\% reduction in uncertainty regarding the specific \ic associated with achieving this cell death probability.
In summary, this result implies that it is optimal to acquire and calibrate to experimental data regarding mRNA-Bax concentrations if we wish to maximize model prediction confidence in which inhibitor dosage achieves a given expected lethality. 

When combining these two metrics to determine an optimal experimental design (i.e., which protein to measure in the lab to reduce uncertainty in lethality curve predictions) we provide an optimal design $\xi^{\ast}$ against the following objectives: (1) $\sigma_{IC_{50}}$ at a threshold of $0.90$, (2) $\sigma_{apop}$ at \ic = 0.01, and (3) weighted averages of both. 
In the case of the weighted averages, we take the objective to be defined as: $w_1 \sigma_{IC_{50}}  + w_2 \sigma_{apop}$  wherein $w_1 + w_2 = 1$. 
For the purposes of illustrating the sensitivity of the combined objective to the selection of weights $w_1, w_2$, we show this weighted average objective under $w_1 = [0.01, 0.1]$. 
These results are tabulated in Table~S\ref{table:objective_results}. 
The results of the weighted average show that at $w_1 = 0.01$, Casp-act is optimal, while at $w_1 = 0.1$, mRNA-Bax is preferred. 
Although the weights used here are arbitrary, domain knowledge should be integrated in the assignment of weights, leading to optimal experiments that align with what is understood regarding the underlying biological pathways. 

In each of the objectives considered, activated caspase or Bax mRNA are identified as the optimal experimental measurements that could maximally reduce model prediction uncertainty if used in calibration.
Here, we provide some insight as to why this may be the case given the present study. 
A possible explanation for the switch from $\xi^{\ast} = $ Casp-act to $\xi^{\ast} = $ mRNA-Bax at low to intermediate \ic may be because of the time dynamics. 
If a higher \ic is needed to kill a cancer cell via apoptosis, then measuring a species that represent the precursor steps to triggering apoptosis may more informative than the activated caspase itself. 
It is also possible that mRNA-Bax performs best across a wider range of drug dosages because it is more proximal to the mechanism of action. 
Once \textit{PARP} is inhibited, only a few reactions are needed to trigger mRNA-Bax formation, and ultimately, the activation of caspase.

These differing outcomes across the different experimental design objectives underscore the essential role of biological insights in differentiating among optimal experimental designs.

\section{Conclusions}\label{sec:conclusions}

In this work, we utilized a realistic PD model for \textit{PARP1}-inhibited cell apoptosis in a model-guided optimal experimental design workflow to reduce uncertainty in the performance of cancer drug candidates. 
We first showed that, under prior uncertainty, the ability for the model to discriminate between \textit{PARP1}-inhibitor therapeutic performance disappears.
Through extensive simulations and Bayesian inference, we have demonstrated the potential of our methodology to reduce parameter uncertainties and reestablish a discriminatory PD model for drug discovery. 
Additionally, our exploration of the \textit{PARP1} inhibited cell apoptosis model has yielded novel insights into species-specific dependencies with \textit{PARP1} inhibitor \ic, informing optimal measurement choices. 
Introducing therapeutic performance metrics based on the half maximal inhibitory concentration (\ic), we provide a comprehensive evaluation of drug efficacy under parameter uncertainty. 
Specifically, we identified optimal experiments under three distinct objectives: (1) $\sigma_{IC_{50}}$ at a threshold of 0.90, (2) $\sigma_{apop}$ at IC50=0.01, and (3) a weighted average of both metrics. 
At low \ic, our results indicate that measuring activated caspases is optimal, with the potential to significantly reduce uncertainty in model-predicted cell death probability.
And at high \ic, Bax mRNA emerges as the favored contender for experimental measurement. 
We note that our algorithmic approach, which includes collecting a large number of posterior samples and then rejecting parameter samples that violate biological constraints, leads to a majority of samples being rejected.
Therefore, in future work, we will explore alternatives for enforcing biological constraints in Bayesian inference, such as relaxed constraints in the data likelihood. 
Additionally, we would like to identify lab experiments that can be conducted to acquire data which is compatible with our model so that we can validate the results of our OED and further constrain the model parameters. 
In summary, our combined metric-based approach not only elucidates the importance of tailored species measurements to minimize uncertainty in lethality curve predictions but also introduces a flexible framework for optimal experimental design using novel biological pathway models. 
Thus, our contributions bridge the critical gap between model-guided optimal design and uncertainty quantification, advancing the realm of computational drug discovery. 

\section{Appendix}\label{sec:STAR}

\subsection{Simulated Experimental Data} \label{STAR:data}

\begin{figure}[H]
    \centering
    \includegraphics[width=0.5\linewidth]{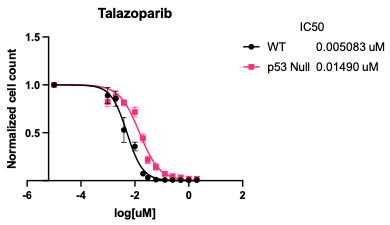}
    \caption{The HCT116 dose response curves. Details on the study can be found in \cite{Mertins2023.02.08.527527}.}
    \label{fig:talazoparib}
\end{figure}

Simulated experimental data was generated for each of the $N_{\xi}$ observable proteins by taking random parameter samples from their respective prior predictive distributions (see Figure S\ref{fig:prior_predictive}) and modifying these samples via a relative error of $10\%$ across all parameters. 
The value of $10\%$ was selected as it is 
the typical standard deviations found in experimental measurements in the laboratory.
First, $N_d$ random draws from the parameter prior distributions are made independently for each uncertain parameter, i.e. $\theta^{sim}_i \sim p(\theta) \quad \forall i = [1,\ldots,N_d]$. 
Next, the $\theta^{sim}_i$ are used to solve the forward model in Equation~S\ref{eqn:ode_model_full} at an \ic=0.005083 $\mu M$, which is the measured \ic for WT cells under the \textit{PARP1}-inhibitor Talazoparib, as shown in Figure~\ref{fig:talazoparib}. 
This generates a ``true" $y$.

The noisy $y^{sim}_{k,i} \quad \forall k = [1,\ldots,N_{\xi}], \forall i = [1,\ldots,N_d]$ used in inference in this work is generated by applying a multiplicative error $\epsilon$, where we assume that the error is log-normally distributed. 
Therefore,

\begin{alignat*}{2}
    & y^{sim}_{k,i} = y_{k,i} \epsilon, & \quad \mathop{log}(\epsilon) \sim \mathcal{N}(0, \sigma^2) \quad \forall k = [1,\ldots,N_{\xi}], \forall i = [1,\ldots,N_d]
\end{alignat*}

Where the standard deviation $\sigma$ is 0.1 for all species $k$. It follows that: 

\begin{alignat*}{2}
    & \mathop{log}(y^{sim}_{k,i}) = \mathop{log}(y_{k,i}) + \mathop{log}(\epsilon) & \\
    & y^{sim}_{k,i} \sim \mathop{exp}(\mathop{log}(y_{k,i}) + \mathop{log(\epsilon)}),
\end{alignat*}

$\forall k = [1,\ldots,N_{\xi}], \forall i = [1,\ldots,N_d]$. To avoid numerical issues at 0 concentrations (i.e., $y_{k,i} = 0$), we employ a parameter $a$ defined such that $a = b - \mathop{min}\{y_{k,i}\}$, where $b$ is a small tolerance (in this study, $b=0.001$). 
Leading to the final form of our simulated experimental data as:

\begin{alignat*}{2}
    & y^{sim} \sim \mathop{exp}(\mathop{log}(y + a) + log(\epsilon)), & \quad \mathop{log}(\epsilon) \sim \mathcal{N}(0, \sigma^2)
\end{alignat*}

\subsection{Algorithmic approach}\label{STAR:algorithm}

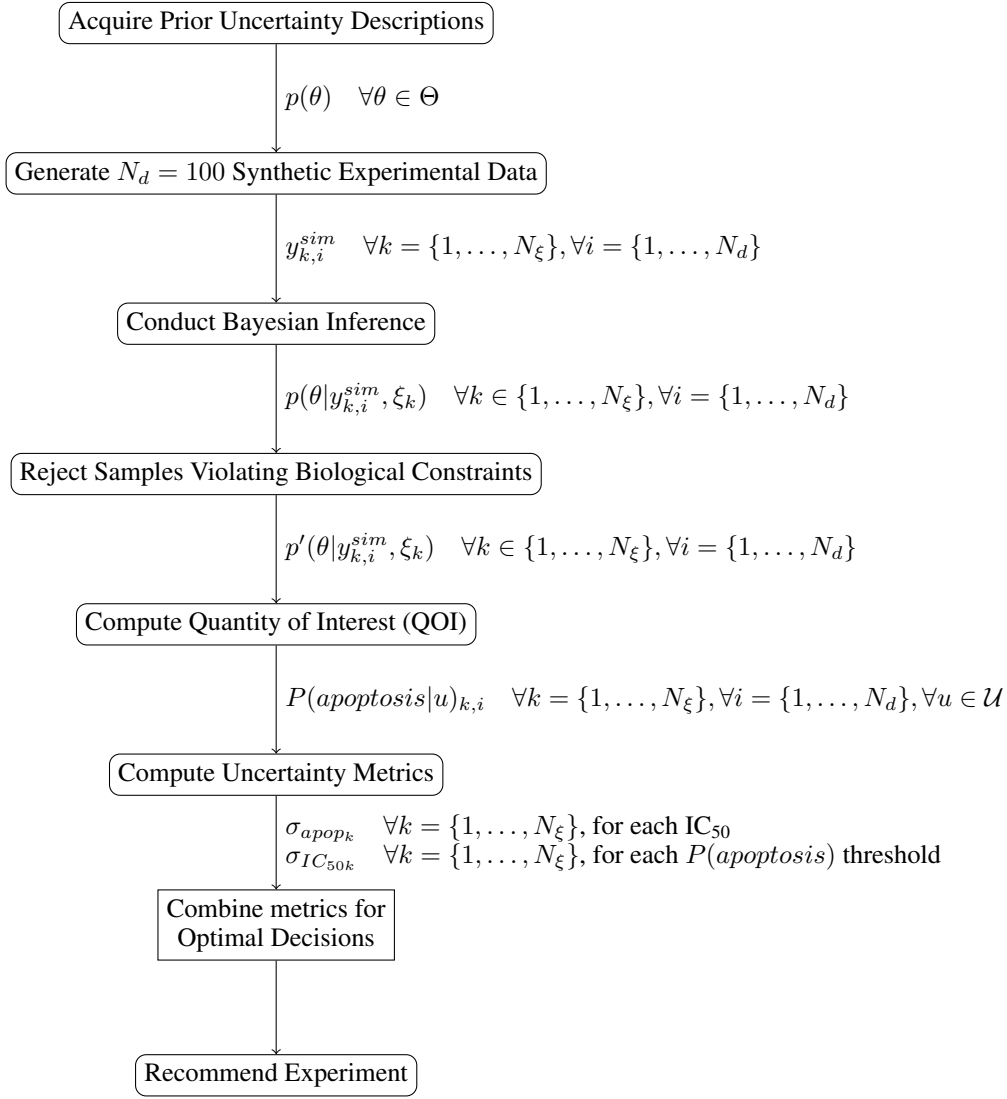
\begin{figure}[H]
    \centering
    \usetikzlibrary{shapes.geometric, arrows}

    \begin{tikzpicture}[node distance=2cm]
        \node (prior) [rectangle, rounded corners, draw] {Acquire Prior Uncertainty Descriptions};
        \node (synthetic) [rectangle, rounded corners, draw, below of=prior] {Generate $N_d=100$ Synthetic Experimental Data};
        \node (inference) [rectangle, rounded corners, draw, below of=synthetic] {Conduct Bayesian Inference};
        \node (constraints) [rectangle, rounded corners, draw, below of=inference] {Reject Samples Violating Biological Constraints};
        \node (qoi) [rectangle, rounded corners, draw, below of=constraints] {Compute Quantity of Interest (QOI)};
        \node (voi) [rectangle, rounded corners, draw, below of=qoi] {Compute Uncertainty Metrics};
        \node (combine) [rectangle, aspect=2, draw, below of=voi, align=center] {Combine metrics for \\ Optimal Decisions};
        \node (recommend) [rectangle, rounded corners, draw, below of=combine] {Recommend Experiment};

        \draw [->] (prior) -- node[right] {$p(\theta) \quad \forall \theta \in \Theta$}(synthetic);
        \draw [->] (synthetic) -- node[right] {$y_{k,i}^{sim} \quad \forall k = \{1,\ldots,N_{\xi}\}, \forall i = \{1,\ldots,N_d\}$}(inference);
        \draw [->] (inference) -- node[right] {$p(\theta|y_{k,i}^{sim}, \xi_k) \quad \forall k \in \{1,\ldots,N_{\xi}\}, \forall i = \{1,\ldots,N_d\}$}(constraints);
        \draw [->] (constraints) -- node[right] {$p'(\theta|y_{k,i}^{sim}, \xi_k) \quad \forall k \in \{1,\ldots,N_{\xi}\}, \forall i = \{1,\ldots,N_d\}$}(qoi);
        \draw [->] (qoi) -- node[right] {$P(apoptosis|u)_{k,i} \quad \forall k = \{1,\ldots,N_{\xi}\}, \forall i = \{1,\ldots,N_d\}, \forall u \in \mathcal{U}$}(voi);
        \draw [->] (voi) -- node[midway, right, align=left] {$\sigma_{apop_{k}} \quad \forall k = \{1,\ldots,N_{\xi}\}$, for each IC\textsubscript{50} \\ $\sigma_{IC_{50 k}} \quad \forall k = \{1,\ldots,N_{\xi}\}$, for each $P(apoptosis)$ threshold}(combine);
        \draw [->] (combine) -- (recommend);

    \end{tikzpicture}
    
    \caption{Workflow utilized for Bayesian experimental design for the \textit{PARP1}-inhibited cell apoptosis model.}
    \label{fig:flowchart}
\end{figure}

The workflow diagram depicted in Figure~\ref{fig:flowchart} outlines the systematic procedure employed for conducting Bayesian Optimal Experimental Design (BOED) within the context of the \textit{PARP1}-inhibited cell apoptosis model. 

\begin{figure}[H]
\centering
     \begin{subfigure}[b]{0.3\textwidth}
         \centering
         \includegraphics[width=\textwidth]{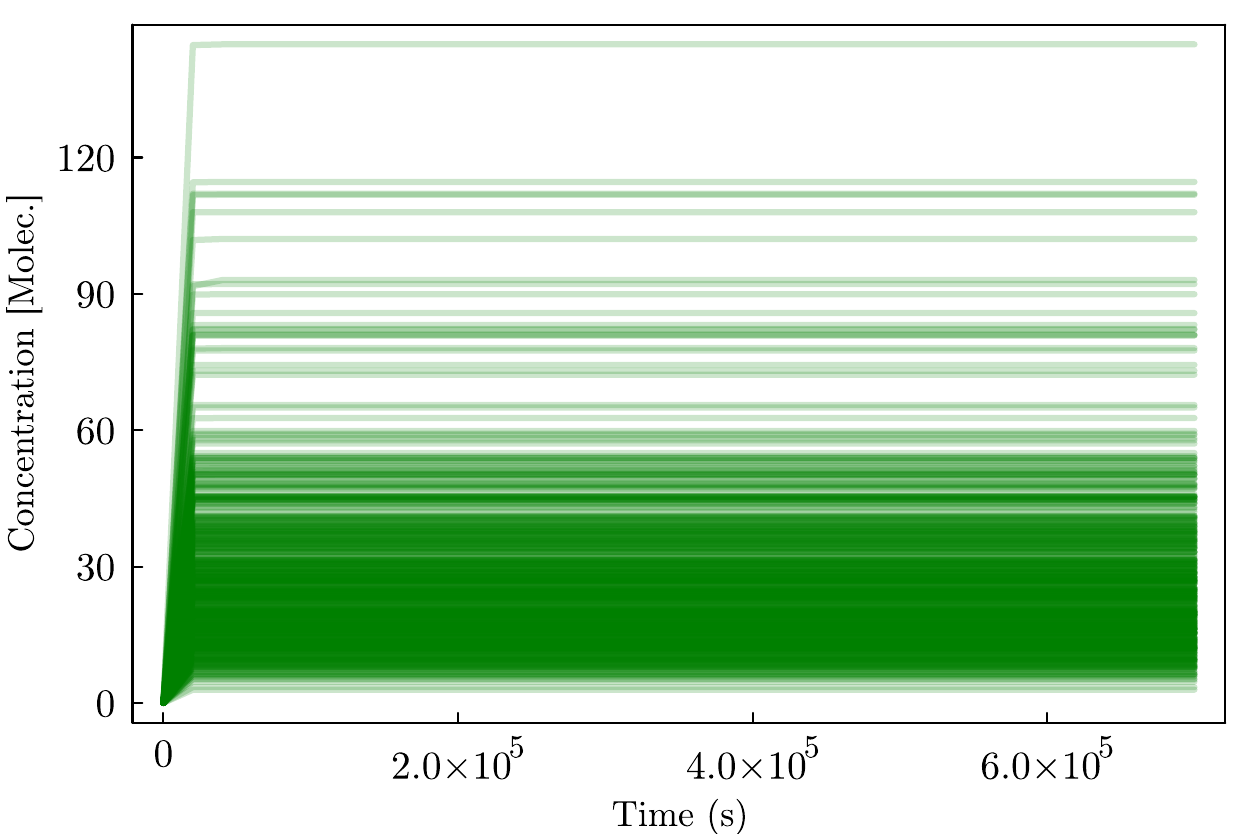}
        \caption{mRNA-Bax}
        \label{fig:prior_pred_13}
     \end{subfigure}
     \hfill
     \centering
     \begin{subfigure}[b]{0.3\textwidth}
         \centering
         \includegraphics[width=\textwidth]{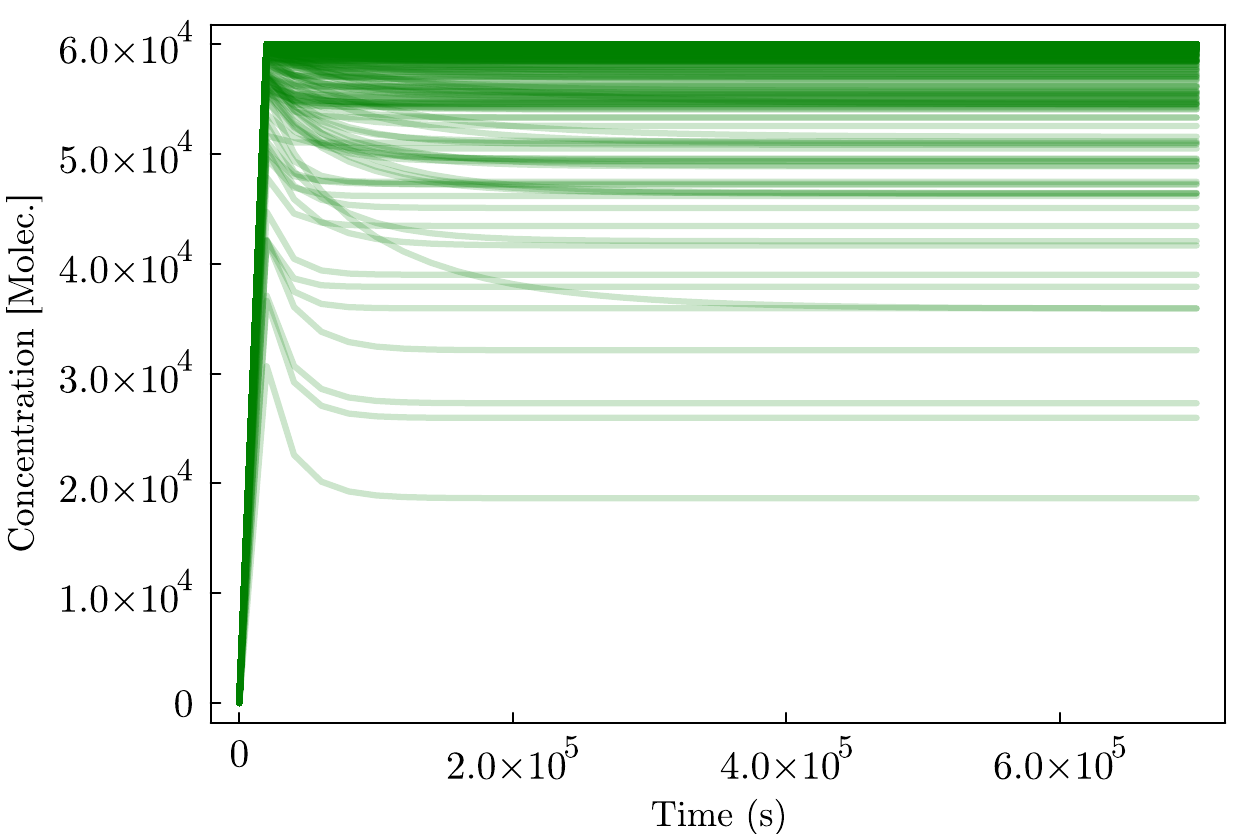}
        \caption{Bad-Bcl-xL}
        \label{fig:prior_pred_15}
     \end{subfigure}
     \hfill
     \centering
     \begin{subfigure}[b]{0.3\textwidth}
         \centering
         \includegraphics[width=\textwidth]{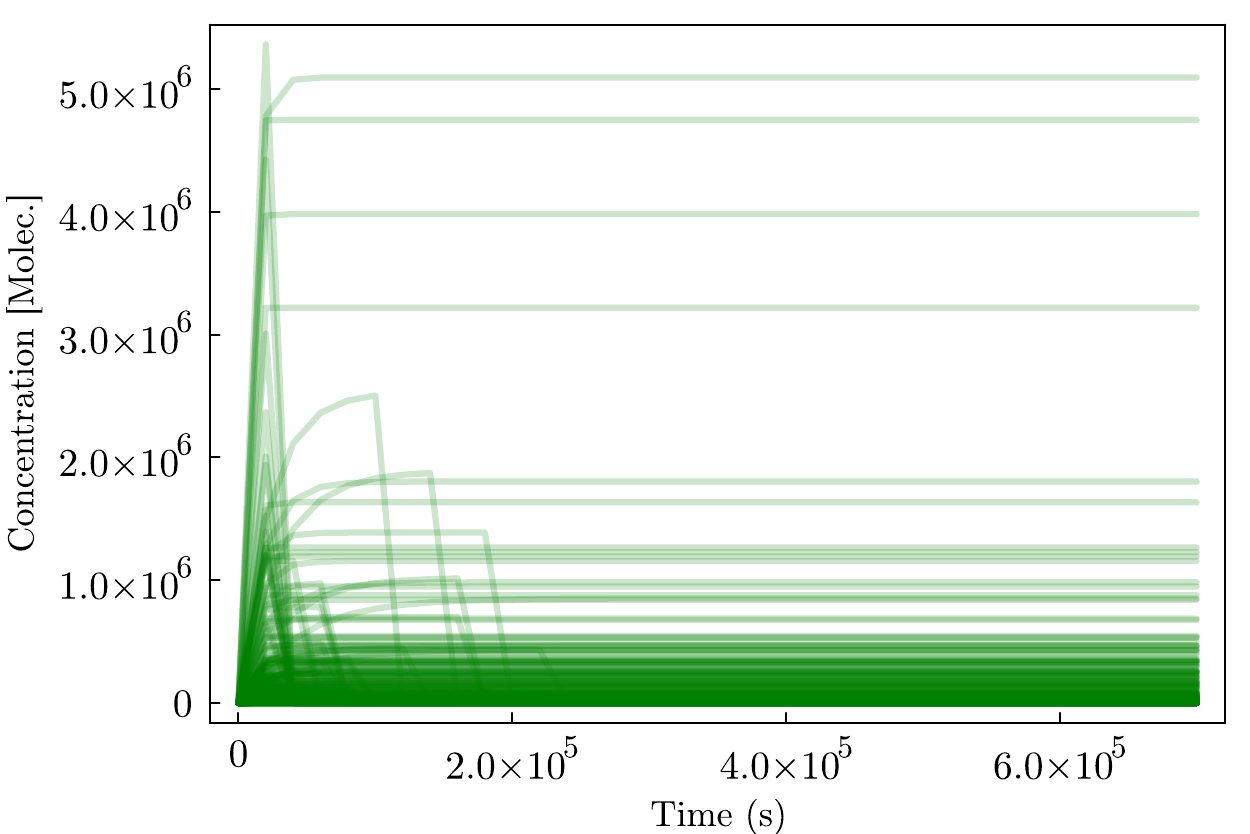}
        \caption{Casp-pro}
        \label{fig:prior_pred_17}
     \end{subfigure}
     \hfill
     \centering
     \begin{subfigure}[b]{0.3\textwidth}
         \centering
         \includegraphics[width=\textwidth]{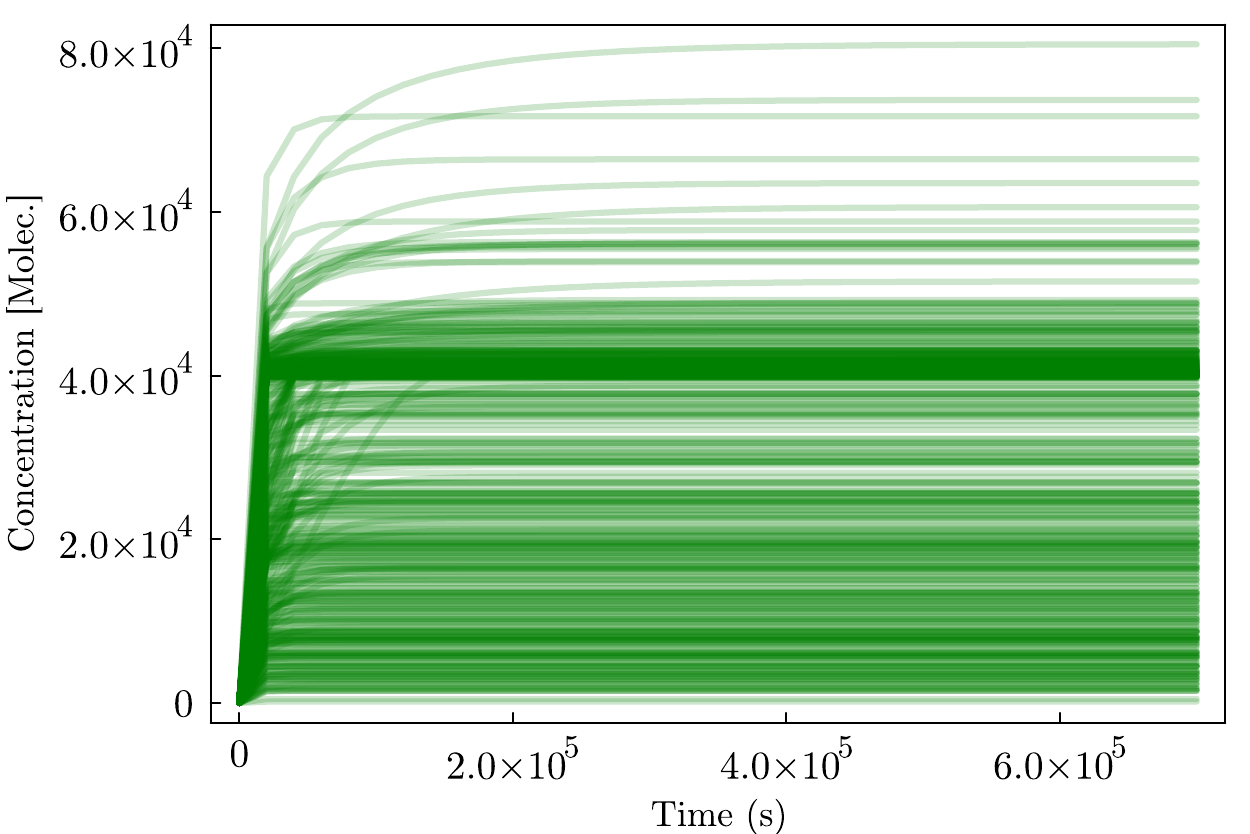}
        \caption{Bax-Bcl-xL}
        \label{fig:prior_pred_20}
     \end{subfigure}
     \hfill
     \centering
     \begin{subfigure}[b]{0.3\textwidth}
         \centering
         \includegraphics[width=\textwidth]{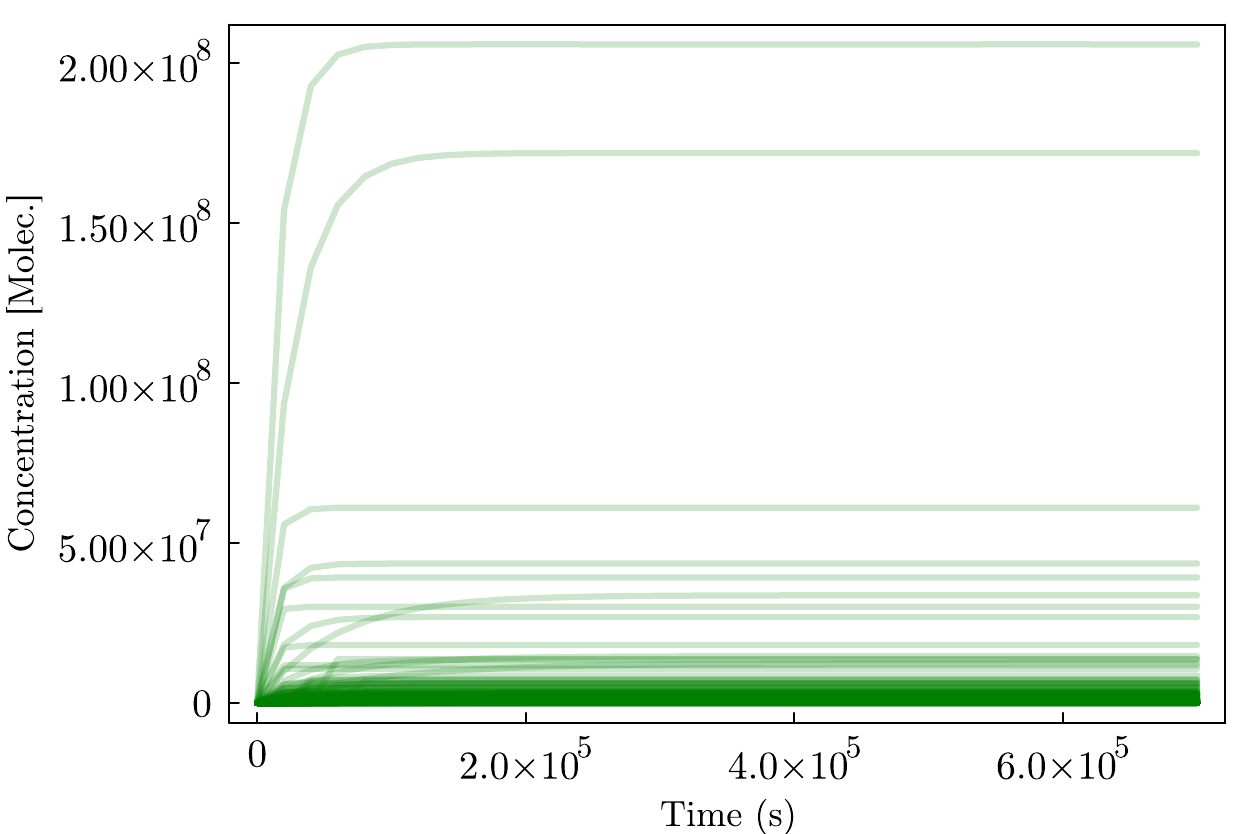}
        \caption{Casp-act}
        \label{fig:prior_pred_22}
     \end{subfigure}
     \caption{Prior predictive plots from 500 simulations sampled randomly from the prior distributions of the uncertain parameters and run through the forward ODE model to create concentration profile distributions. Units of concentration are in molecules-per-cell.}\label{fig:prior_predictive}
\end{figure}

The process begins by specifying prior uncertainty descriptions for the key model parameters ($p(\theta) \ \forall \theta \in \Theta$) determined by global sensitivity analysis results presented in \cite{Mertins2023.02.08.527527}. 
This subset is shown in Table~S\ref{table:priors}.
Next, synthetic experimental data $(y^{sim}_{k,i})$ are generated for each experimental design (indexed by $k = \{1,\ldots,N_{\xi}\}$) under consideration, as listed in Table~S\ref{table:observables}, and for $i = \{1,\ldots,N_d\}$ samples from the prior distributions and a known error model (see Table~S\ref{appendix:distributions}). 
We note that we select the experimental designs listed in Table~S\ref{table:observables} because these species exhibited high prior predictive uncertainty, as seen in Figure~S\ref{fig:prior_predictive}. 
These synthesized data are used to constrain the uncertain model parameters via Bayesian Inference, where the posterior parameter probability distributions ($p(\theta|y_{k,i}^{sim}, \xi_k)$) are estimated using a HMC algorithm (i.e., a type of MCMC algorithm for sampling from an unknown distribution) for each $k = \{1,\ldots,N_{\xi}\}$ and for each datum $i = \{1,\ldots,N_d\}$. 
Next, constraints based on biological feasibility are then applied to reject samples in the Markov chains that violate the expected behavior of the cell under extreme \textit{PARP1} inhibitor dosages. 
This leads to modified posterior estimates ($p'(\theta|y_{k,i}^{sim}, \xi_k)$). 
These posterior distributions are then utilized to compute the Quantity of Interest (QOI), specifically the probability of cell apoptosis ($P(apoptosis|u)$), which is assessed for various inhibitor \ic values ($u \in \mathcal{U}$). 
These QOI values serve as the basis for computing the uncertainty metrics: $\sigma_{apop}(u)$ and $\sigma_{IC_{50 k}}(\mathbb{T})$ for each experimental design. 
Finally, the uncertainty metrics are combined to make optimal experimental decisions, thereby providing recommendations for the most effective experimental designs. 
This comprehensive workflow guides the process of leveraging BOED to enhance our understanding of how we can optimally reduce uncertainty in the model predictions made by the \textit{PARP1}-inhibited cell apoptosis model under different drug dosage conditions.

\subsection{Uncertainty Metrics}\label{STAR:voi_metrics}

The calculated lethality serves as the basis for deriving our fundamental experimental design objectives: the $\sigma_{IC_{50}}(\mathbb{T})$ and $\sigma_{apop}(u)$ uncertainty metrics. These metrics offer a quantifiable means of assessing the reduction in uncertainty associated with the various experimental designs, $\xi$.

The $\sigma_{IC_{50}}(\mathbb{T})$ metric is determined by fitting a 4-parameter sigmoidal curve, specifically the form $P(apoptosis|IC_{50})_i = d_i + \frac{a_i-d_i}{1+(\frac{IC{50}}{c_i})^{b_i}}$, to the computed $P(apoptosis|u)_i$ values for each \ic value and for each $i \in [1,\ldots,N_d]$. Subsequently, the intersection points of the sigmoidal fits with pre-specified $P(apoptosis)$ thresholds (e.g., $\mathbb{T}$ = [0.5, 0.6, 0.7, 0.8, 0.9]) determine the associated \ic values. The $\sigma{IC_{50}}(\mathbb{T})$ metric is then defined as the standard deviation over \ic values at which the sigmoidal fits intersect the threshold.

The $\sigma_{apop}$ metric is computed by taking the standard deviation over all predicted probabilities of apoptosis at a given \ic, i.e., $P(apoptosis|u)_i, \quad \forall i = [1,\ldots,N_d], \forall u \in \mathcal{U}$. This is defined in Equation~\ref{eqn:def_sig_apop}, where $\bar{P}(apoptosis|u)$ is the arithmetic mean of the predicted probabilities of cell death at the specified inhibitor \ic, or $u$.

\begin{alignat}{1}
    & \displaystyle \sigma_{apop}(u) := \sqrt{\frac{1}{N_{d} - 1} \sum_{i=1}^{N_d}  \left(P(apoptosis|u) - \bar{P}(apoptosis|u)\right)} \quad \forall u \in \mathcal{U}
    \label{eqn:def_sig_apop}
\end{alignat}

\subsection{Implementation}

The entirety of the workflow presented here was implemented using the programming language Julia \cite{bezanson2012julia}.
Specifically, we utilize the probabilistic programming library Turing.jl\cite{ge2018turing} to implement our probabilistic model for Bayesian inference. 
To sample the posterior chain and obtain posterior probability distributions, we use the No U-Turn Sampler (NUTS), a variant of HMC, with an acceptance rate of 0.65.
The ODE model evaluations within the probabilistic model are conducted via the DifferentialEquations.jl library \cite{rackauckas2017differentialequations}. 

The implementation of BOED we propose here exploits several elements of high-performance computing approaches to accelerate the optimization process. 
First, the implementation of the ODE model was translated into the low-level programming language, Julia, from the rule-based modeling language, BioNetGen \cite{faeder2009rule}. 
When performing ODE model simulations at the nominal parameter settings, the BioNetGen implementation requires 0.10 seconds of CPU time. 
In contrast, solving the ODE model under the same nominal parameter settings using Julia takes only 0.0018 seconds of CPU time. 
Consequently, our Julia-based model implementation exhibits an impressive 98\% reduction in CPU time compared to the BioNetGen implementation. 
Although these individual execution times may already appear low, the acceleration of individual model evaluations is critically important in the context of Bayesian optimal experimental design. 
This is because these model evaluations are used in estimating the data likelihood across various parameter values, and as the number of these evaluations increases, the cumulative effect on computational time of Bayesian inference becomes more pronounced. 
Secondly, we accelerate the computation of each of the 100 posterior estimates per experimental design by using 100 CPU cores in parallel.
This means that we can compute 100 posteriors simultaneously, decreasing the time-to-solution over a sequential approach. 
Thus, we show that the combined utilization of these high-performance computing strategies enhances the efficiency and practical applicability of our proposed Bayesian optimal experimental design framework within the context of a complex PD model. 
All instances of posterior sampling through NUTS were solved on the Institutional Cluster at the Scientific Data and Computing Center at Brookhaven National Library. Runs were done in parallel using 3 nodes at a time. Each node used consists of 2 CPUs Intel(R) Xeon(R) CPU E5-2695 v4 @ 2.10GHz with 18 cores each.

\section*{Funding}\label{sec:funding}
\begin{flushleft}
This work was supported by the New York Empire State Development Big Data 800 Science Capital Project, the DOE Advanced Scientific Computing Research Applied 801 Mathematics Fellowship, and the Brookhaven Laboratory Directed Research and 802 Development DEDUCE Project. (NMI, KRR, BJY, NMU)

Funding sources include federal funds from the National Cancer Institute, National 793 Institutes of Health, and the Department of Health and Human Services, Leidos Biomedical 794 Research Contract No. 75N91019D00024, Task Order 75N91019F00134 through the 795 Accelerating Therapeutics for Opportunities in Medicine (ATOM) Consortium under 796 CRADA TC02349. (SDM)
\end{flushleft}

\section*{Acknowledgements}\label{sec:acknowledgements}

\begin{flushleft}
The authors would like to acknowledge Manasi P. Jogalekar and Morgan E. Diolaiti for providing experimental data related to the \ic for Talazoparib used in this work. 

In addition, we acknowledge the roles of Naomi Ohashi and Eric A. Stahlberg for their project management roles at the Frederick
National Laboratory for Cancer Research and ATOM Research Alliance.
\end{flushleft}

\section*{Author Contributions}\label{sec:contrib}

Conceptualization of problem, NMI, SDM, BJY, KR, and NMU; Methodology, NMI, SDM, BJY, KR, and NMU; Implementation, NMI; Analysis, NMI, SDM, BJY, KR, and NMU; Writing -- original draft, NMI; Writing -- reviewing and editing, NMI, SDM, BJY, KR, and NMU.

\section*{Declaration of Interests}\label{sec:declarations}

SDM is Founder and CEO of BioSystems Strategies, LLC.

NMI, BJY, KR, and NMU declare no competing interests.


\section*{Supplemental Information}

\section{Uncertain Parameter Information} \label{appendix:distributions}

Prior information for the uncertain parameters considered in this study was first elicited from expert understanding of the system and relevant sources and is compiled in Table~\ref{table:priors}. 
This table shows the model parameters, their nominal values, and estimated 90\% confidence interval lower and upper limits.

\begin{table}[H]
\centering
\begin{tabular}{||c l l l l||} 
 \hline
 Parameter & Units & Nominal Value & Lower Confidence Bound & Upper Confidence Bound \\ [0.5ex] 
 \hline\hline
 $s_1$ & mol s$^{-1}$ & \num{0.01} & \num{0.0045} & \num{0.022} \\ 
 $s_2$ & mol s$^{-1}$ & \num{3e-2} & \num{0.013} & \num{0.067} \\
 $s_3$ & mol s$^{-1}$ & \num{2e1} & \num{0.42} & \num{942} \\
 $s_4$ & mol s$^{-1}$ & \num{2e-1} & \num{0.045} & \num{0.89} \\
 $d_1$ & s$^{-1}$ & \num{1e-3} & \num{0.00045} & \num{0.0022} \\  
 $d_2$ & s$^{-1}$ & \num{1e-4} & \num{0.000020} & \num{0.00051}  \\  
 $d_3$ & s$^{-1}$ & \num{2e-4} & \num{0.000039} & \num{0.0010} \\  
 $M$ & - & 100,000 & {90,453} & {110,554} \\  
 $b_1$ & mol$^{-1}$ s$^{-1}$ & \num{3e-5} & \num{0.0000039} & \num{0.00023} \\  
 $a_1$ & mol$^{-1}$ s$^{-1}$ & \num{2e-10} & \num{4.14e-11} & \num{9.66e-10}  \\  
 $a_2$ & mol$^{-1}$ s$^{-1}$ &\num{1e-12} & \num{6.55e-13} & \num{1.53e-12} \\
 \hline 
\end{tabular}
\caption{Table of uncertain parameters and 90\% confidence limits for the \textit{PARP1} inhibitor ODE model. All prior probability distributions are assumed to be log-normal to prohibit negative rate parameter values.}
\label{table:priors}
\end{table}

Derived prior probability distributions for each uncertain parameter considered in the model are described in Table~\ref{table:prior_distributions}. All priors are defined as Log-Normal probability distributions. For all parameters, the nominal value was taken as the mean. Because the lower and upper bounds on the uncertain parameters are not centered at the nominal value, the width of this CI range was used to compute the geometric distance from the nominal to new lower and upper bounds in log-space. 

\begin{table}[H]
\centering
\begin{tabular}{||c l l ||} 
 \hline
 Parameter & Log Mean & Log Std \\ [0.5ex] 
 \hline\hline
 $s_1$ & -4.61 & 0.49 \\ 
 $s_2$ & -3.51 & 0.49 \\ 
 $s_3$ & 3.00 & 2.34 \\ 
 $s_4$  & -1.61 & 0.91 \\ 
 $d_1$ & -6.91 & 0.49\\   
 $d_2$  & -9.21 & 1.00 \\   
 $d_3$  & -8.52 & 1.00 \\   
 $M$ & 11.51 & 0.06 \\   
 $b_1$  & -10.41 & 1.24 \\   
 $a_1$  & -22.33 & 0.96 \\   
 $a_2$  & -27.63 & 0.26 \\ 
 \hline
\end{tabular}
\caption{Table of distributional information (mean, standard deviation) for Log-Normal priors on all uncertain parameters considered in this work.}
\label{table:prior_distributions}
\end{table}

\section{Experimental Designs}\label{appendix:observables}

The list of measurable species considered as potential ``experimental designs'' in the present study are shown in Table~\ref{table:observables}.

\begin{table}[H]
\centering
\begin{tabular}{||c l||} 
 \hline
 Species Symbol & Definition  \\ [0.5ex] 
 \hline\hline
 mRNA-Bax & Messenger mRNA for Bax  \\  
 Bad-Bcl-xL & Bad bound to Bcl-xL   \\ 
 Casp-pro & Inactive caspase   \\
 Bax-Bcl-xL & Bax bound to Bcl-xL \\
 Casp-act & Activated caspase  \\
 \hline
\end{tabular}
\caption{Table of measurable species in the \textit{PARP1} inhibited ODE model. As in \cite{Mertins2023.02.08.527527}, all initial values for these species is 0. The ODE model computes concentrations in dimensionless units \textit{[Molec.]} representing molecules-per-cell.}
\label{table:observables}
\end{table}

\section{Tabulated Objective Data}\label{appendix:objective_data}

\begin{table}[ht]
\centering
\begin{tabular}{||c|c|c|c|c|c|c|c|c|c|c||} 
 \hline
 \multirow{2}{*}{Measured Protein}  & \multicolumn{10}{c||}{$\sigma_{apop}$} \\
        \cline{2-11} 
      & $IC_{50}=$ & 0.001 & 0.01 & 0.03 & 0.1 & 0.3 & 1.0 & 3.0 & 10 & 100 \\
         \hline\hline
 mRNA-Bax & & 0.0 & {0.00446}	& {0.0077}	& {0.0128}$^{\ast}$ & {0.0190}$^{\ast}$ & {0.0261}$^{\ast}$ & {0.0231}$^{\ast}$	& {0.0190}$^{\ast}$ & 0.0\\ 
 Bad-Bcl-xL & & 0.0 & 0.00446 & 0.0076 & 0.0144 & 0.0199 & 0.0277 & 0.0271 & 0.0195 & 0.0\\ 
 Casp-pro & & 0.0 & 0.0119 & 0.0205 & 0.0355 & 0.0413 & 0.0494 & 0.0435 & 0.0273 & 0.0 \\ 
 Bax-Bcl-xL & & 0.0 & 0.0041 & 0.0095 & 0.0342 &  0.0353 & 0.0309 & 0.0283 & 0.0211 & 0.0  \\ 
 Casp-act & & 0.0 & 0.0034$^{\ast}$ & 0.0074$^{\ast}$ & 0.0158 & 0.0289 & 0.0421 & 0.0418 & 0.0277 & 0.0 \\ 
 \hline
\end{tabular}
\caption{Table of $\sigma_{apop}$ results for each measureable protein in the \textit{PARP1}-inhibited cell apoptosis model. The minimum value in each column is denoted with an $^{\ast}$.}
\label{table:sigma_apop}
\end{table}

\begin{table}[ht]
\centering
\begin{tabular}{||c|c|c|c|c|c|c||} 
 \hline
 \multirow{2}{*}{Measured Species}  & \multicolumn{6}{c||}{$\sigma_{IC_{50}}$} \\
        \cline{2-7} 
      & Threshold= & 0.5 & 0.6 & 0.7 & 0.8 & 0.9 \\
         \hline\hline
 mRNA-Bax &  & 0.170$^{\ast}$ & 0.113$^{\ast}$ & 0.075$^{\ast}$ & 0.047$^{\ast}$ & 0.023$^{\ast}$ \\
 Bad-Bcl-xL &  & 0.179 & 0.117 & 0.078 & 0.049 & 0.026 \\
 Casp-pro &  & 0.327 & 0.226 & 0.153 & 0.095 & 0.047  \\ 
 Bax-Bcl-xL &  & 0.208 & 0.147 & 0.102 & 0.066 & 0.033\\ 
 Casp-act &  & 0.390 & 0.278 & 0.179 & 0.109 & 0.053 \\ 
 \hline
\end{tabular}
\caption{Table of $\sigma_{IC_{50}}$ results for each measureable protein in the \textit{PARP1}-inhibited cell apoptosis model. The minimum value in each column is denoted with an $^{\ast}$.}
\label{table:sigma_ic50}
\end{table}

\begin{table}[H]
\centering
\begin{tabular}{||c c c c||} 
 \hline
 $\sigma_{apop}(0.01)$ & $\sigma_{IC_{50}}(0.9)$ & $w_1$ = 0.01 & $w_1$ = 0.1\\ [0.5ex] 
 \hline\hline
 $\xi^{\ast}$ = Casp-act &  $\xi^{\ast}$ = mRNA-Bax & $\xi^{\ast}$ = Casp-act 
 & $\xi^{\ast}$ = mRNA-Bax \\
 \hline
\end{tabular}
\caption{Table of optimal experimental designs, $\xi^{\ast}$ for each objective considered in the present study.}
\label{table:objective_results}
\end{table}

\section{Posterior Chain Summary Statistics}\label{appendix:chain_data}

Each table in this section presents summary statistics for uncertain parameter values and effective sample size for each uncertain parameter inferred in Bayesian inference against a specific species measurement. 
Averages are computed over 100 synthetic experimental datum for that protein generated according to the procedure described in the STAR Methods. An additional table is provided per each species to list the min, max, and average chain size from all 100 posteriors used in obtaining these statistics.

\subsection*{mRNA-Bax}

\begin{table}[H]
\centering
\begin{tabular}{||c c c||} 
 \hline
 Parameter & Mean Value & Average ESS  \\ [0.5ex] 
 \hline\hline
    $s_1$ & \num{0.0101} & 561 \\
    $s_2$ & \num{0.0370} & 565\\
    $s_3$ & 141.3 & 566 \\
    $s_4$ & \num{0.280} & 563 \\
    $d_1$ & \num{0.00114} & 560 \\
    $d_2$ & \num{0.000133} & 550 \\
    $d_3$ & \num{0.000274} & 569 \\
    $M$ & 100,276 & 565 \\
    $b_1$ & \num{6.82E-05} & 566 \\
    $a_1$ & \num{3.27E-10} & 562 \\
    $a_2$ & \num{1.03E-12} & 549 \\
 \hline
\end{tabular}
\caption{Average expected parameter values and effective sample size from filtered chains for posteriors obtained via mRNA-Bax data.}\label{table:chain_data_mRNA_Bax}
\end{table}

\begin{table}[H]
\centering
\begin{tabular}{||c c c||} 
 \hline
 Min & Max & Average \\ [0.5ex] 
 \hline\hline
    348 & 690 & 594 \\
    \hline
\end{tabular}
\caption{Chain length statistics (min, max, and average) after sample rejection for mRNA-Bax.}\label{table:stat_mrna}
\end{table}

\subsection*{Bad-Bcl-xL}

\begin{table}[H]
\centering
\begin{tabular}{||c c c||} 
 \hline
 Parameter & Mean Value & Average ESS  \\ [0.5ex] 
 \hline\hline
    $s_1$ & \num{0.0102} & 549 \\
    $s_2$ & \num{0.0374} & 541 \\
    $s_3$ & 136.875 & 542 \\
    $s_4$ & \num{0.282} & 540 \\
    $d_1$ & \num{0.00112} & 545 \\
    $d_2$ & \num{0.000136} & 531 \\
    $d_3$ & \num{0.000279} & 539 \\
    $M$ & 101,239  & 554 \\
    $b_1$ & \num{7.09E-05} & 525 \\
    $a_1$ & \num{3.30E-10} & 550 \\
    $a_2$ & \num{1.04E-12}  & 557 \\
 \hline
\end{tabular}
\caption{Average expected parameter values and effective sample size from filtered chains for posteriors obtained via Bad-Bcl-xL data.}
\end{table}

\begin{table}[H]
\centering
\begin{tabular}{||c c c||} 
 \hline
 Min & Max & Average \\ [0.5ex] 
 \hline\hline
    111 & 715 & 606\\
    \hline
\end{tabular}
\caption{Chain length statistics (min, max, and average) after sample rejection for Bad-Bcl-xL.}\label{table:stat_bad_bcl_xl}
\end{table}

\subsection*{Casp-pro}

\begin{table}[H]
\centering
\begin{tabular}{||c c c||} 
 \hline
 Parameter & Mean Value & Average ESS  \\ [0.5ex] 
 \hline\hline
    $s_1$ & \num{7.91E-03}  & 226 \\
    $s_2$ & \num{0.0284} & 238 \\
    $s_3$ & 34.15  & 195 \\
    $s_4$ & \num{0.24} & 235 \\
    $d_1$ & \num{7.93E-04}  & 233 \\
    $d_2$ & \num{8.19E-05}  & 228 \\
    $d_3$ & \num{3.07E-04} & 195 \\
    $M$ & 75,191  & 224 \\
    $b_1$ & \num{4.90E-05}  & 228 \\
    $a_1$ & \num{2.47E-10} & 219 \\
    $a_2$ & \num{7.50E-13}  & 220 \\
 \hline
\end{tabular}
\caption{Average expected parameter values and effective sample size from filtered chains for posteriors obtained via Casp-pro data.}
\end{table}

\begin{table}[H]
\centering
\begin{tabular}{||c c c||} 
 \hline
 Min & Max & Average \\ [0.5ex] 
 \hline\hline
    192 & 1343 & 648 \\
    \hline
\end{tabular}
\caption{Chain length statistics (min, max, and average) after sample rejection for Casp-pro.}\label{table:stat_casp_pro}
\end{table}

\subsection*{Bax-Bcl-xL}

\begin{table}[H]
\centering
\begin{tabular}{||c c c||} 
 \hline
 Parameter & Mean Value & Average ESS  \\ [0.5ex] 
 \hline\hline
    $s_1$ & \num{0.0102} & 623 \\
    $s_2$ & \num{0.0373} & 620 \\
    $s_3$ & 129.6 & 616 \\
    $s_4$ & \num{0.283} & 614 \\
    $d_1$ & \num{0.00110}  & 615 \\
    $d_2$ & \num{0.000132}  & 606 \\
    $d_3$ & \num{0.000282 }& 610 \\
    $M$ & 100,209  & 610 \\
    $b_1$ & \num{7.48E-05} & 606 \\
    $a_1$ & \num{3.25E-10} & 607 \\
    $a_2$ & \num{1.03E-12}  & 614 \\
 \hline
\end{tabular}
\caption{Average expected parameter values and effective sample size from filtered chains for posteriors obtained via Bax-Bcl-xL data.}
\end{table}

\begin{table}[H]
\centering
\begin{tabular}{||c c c||} 
 \hline
 Min & Max & Average \\ [0.5ex] 
 \hline\hline
    107 & 1115 & 812 \\
    \hline
\end{tabular}
\caption{Chain length statistics (min, max, and average) after sample rejection for Bax-Bcl-xL.}
\label{table:stat_bax_bcl_xl}
\end{table}

\subsection*{Casp-act}

\begin{table}[H]
\centering
\begin{tabular}{||c c c||} 
 \hline
 Parameter & Mean Value & Average ESS  \\ [0.5ex] 
 \hline\hline
    $s_1$ & \num{0.00900} & 249 \\
    $s_2$ & \num{0.0331} & 250 \\
    $s_3$ & 110.6 & 243 \\
    $s_4$ & \num{0.211} & 249 \\
    $d_1$ & \num{0.00105} & 245 \\
    $d_2$ & \num{0.000139} & 247 \\
    $d_3$ & \num{0.000180} & 242 \\
    $M$ & 91,317 & 259 \\
    $b_1$ & \num{6.26E-05} & 249 \\
    $a_1$ & \num{2.83E-10} & 247 \\
    $a_2$ & \num{9.51E-13} & 253 \\
 \hline
\end{tabular}
\caption{Average expected parameter values and effective sample size from filtered chains for posteriors obtained via Casp-act data.}
\end{table}

\begin{table}[H]
\centering
\begin{tabular}{||c c c||} 
 \hline
 Min & Max & Average \\ [0.5ex] 
 \hline\hline
    265 & 1568 & 1180 \\
    \hline
\end{tabular}
\caption{Chain length statistics (min, max, and average) after sample rejection for Casp-act.}
\label{table:stat_casp_act}
\end{table}

\section{Prior predictive distributions}\label{appendix:prior_predictive}

\section{Mathematical Model}\label{appendix:ode_model_full}

This section contains the full system of differential equations that define the \textit{PARP1}-inhibited cell apoptosis model as presented in \cite{Mertins2023.02.08.527527}. In the notation in Equation~\ref{eqn:ode_model_full}, $\dot{x}_i$ refers to the time-varying rate of change of species $i$, where $i=\{1,\ldots,23\}$, $x_i$ is the species concentration in units of molec/cell, and $p_i, \ i=\{1,\ldots,27\}$ are the ODE parameters related to rates. See Tables~\ref{table:ode_species} and ~\ref{table:ode_params} for a mapping from $x_i$ and $p_i$ notation to biological symbols and values. 
This mathematical model was generated from the rule-based modeling software, BioNetGen, using the previously published PARP1-inhibited cell apoptosis model in \cite{Mertins2023.02.08.527527}.

\setlength{\abovedisplayskip}{-3pt}
\begin{align}\label{eqn:ode_model_full}
\dot{x_{1}}&=  - p_{12} x_{2} x_{1} - p_{12} x_{6} x_{1} + p_{16} x_{11} + p_{16} x_{12} + p_{24} x_{10} x_{12} - p_{25} x_{19} x_{1} \nonumber
\\
\dot{x_{2}}&=  - p_{12} x_{2} x_{1} + p_{16} x_{11}
\nonumber\\
\dot{x_{3}}&=  - p_{10} x_{3} x_{4} - p_{9} x_{14} x_{3} + p_{14} x_{15} + p_{42} x_{15} + p_{13} x_{20} + p_{6} x_{20}
\nonumber\\
\dot{x_{4}}&=  - p_{10} x_{3} x_{4} - p_{41} x_{4} + p_{14} x_{15} + p_{18} x_{16}+ p_{18} x_{21}
\nonumber\\
\dot{x_{5}}&=  - p_{11} x_{16} x_{5} + p_{15} x_{21} + p_{18} x_{21}
\nonumber\\
\dot{x_{6}}&=  - p_{12} x_{6} x_{1} + p_{16} x_{12}
\nonumber\\
\dot{x_{7}}&=  - p_{21} x_{12} x_{7} + p_{22} x_{18} + p_{26} x_{18} x_{8}
\nonumber\\
\dot{x_{8}}&=  - p_{26} x_{18} x_{8} + p_{27} x_{9} x_{23}
\nonumber\\
\dot{x_{9}}&= 0
\nonumber\\
\dot{x_{10}}&=  - p_{24} x_{10} x_{12} + p_{25} x_{19} x_{1}
\nonumber\\
\dot{x_{11}}&=  + p_{12} x_{2} x_{1} - p_{16} x_{11}
\nonumber\\
\dot{x_{12}}&=  + p_{12} x_{6} x_{1} - p_{16} x_{12} - p_{21} x_{12} x_{7} - p_{24} x_{10} x_{12} + p_{22} x_{18} + p_{26} x_{18} x_{8} + p_{25} x_{19} x_{1}
\\
\dot{x_{13}}&=  + p_{1} + p_{39} - p_{5} x_{13}
\nonumber\\
\dot{x_{14}}&=  + p_{40} - p_{6} x_{14} - p_{9} x_{14} x_{3} + p_{13} x_{20}
\nonumber\\
\dot{x_{15}}&=  + p_{10} x_{3} x_{4} - p_{14} x_{15} - p_{42} x_{15}
\nonumber\\
\dot{x_{16}}&=  + p_{41} x_{4} - p_{18} x_{16} - p_{11} x_{16} x_{5} + p_{42} x_{15} + p_{15} x_{21}
\nonumber\\
\dot{x_{17}}&=  + p_{3} - p_{7} x_{17} - p_{43} x_{17}
\nonumber\\
\dot{x_{18}}&=  + p_{21} x_{12} x_{7} - p_{22} x_{18} - p_{26} x_{18} x_{8}
\nonumber\\
\dot{x_{19}}&=  + p_{24} x_{10} x_{12} - p_{25} x_{19} x_{1}
\nonumber\\
\dot{x_{20}}&=  + p_{9} x_{14} x_{3} - p_{13} x_{20} - p_{6} x_{20}
\nonumber\\
\dot{x_{21}}&=  + p_{11} x_{16} x_{5} - p_{15} x_{21} - p_{18} x_{21}
\nonumber\\
\dot{x_{22}}&=  + p_{43} x_{17} - p_{7} x_{22}
\nonumber\\
\dot{x_{23}}&=  + p_{26} x_{18} x_{8} - p_{27} x_{9} x_{23}
\nonumber
\end{align}

\begin{table}[H]
\centering
\begin{tabular}{||c l l l||} 
 \hline
 Symbol & Species & Description & Initial Values\\ [0.5ex] 
 \hline\hline
  $x_1$ & DNADSB & DNA double strand break & \num{1.74e5}\\
  $x_2$ & p53 & p53 transcription factor & \num{8.5e4}\\
  $x_3$ & Bcl-xL & Anti-apoptotic factor & \num{1e5}\\
  $x_4$ & Bad & Pro-apoptotic factor & \num{6e4}\\
  $x_5$ & Scaffold 14-3-3 & Scaffold 14-3-3 & \num{2e5}\\
  $x_6$ & PARP & Poly(ADP-ribose) polymerase & \num{1.07e5}\\
  $x_7$ & NAD & Nicotinamide adenosine deoxyribose & \num{1.07e6}\\
  $x_8$ & XRCC1(Glu$\sim$uPAR) & PARP substrate XRCC1 unPARylated & \num{1.07e5}\\
  $x_9$ & PARG & Poly(ADP-ribose) glycohydrase & \num{1.07e4}\\
  $x_{10}$ & Inhibitor & Generic PARP inhibitor & 0.005083\\
  $x_{11}$ & DNADSB-p53 & p53 bound to DNA double strand break & 0\\
  $x_{12}$ & DNADSB-PARP & PARP bound to DNA double strand break & 0\\
  $x_{13}$ & mRNA-Bax & Messanger RNA for Bax & 0\\
  $x_{14}$ & Bax & Pro-apoptotic factor & 0\\
  $x_{15}$ & Bad-Bcl-xL & Bad bound to Bcl-xL & 0\\
  $x_{16}$ & Bad(S75 S99$\sim$PP,b) & Phosphorylated Bad& 0\\
  $x_{17}$ & Casp-pro & Inactive caspase & 0\\
  $x_{18}$ & DNADSB-NAD-PARP & NAD bound to PARP active site while PARP bound to DNA & 0\\
  $x_{19}$ & Inh-PARP & Inhibitor bound to PARP & 0\\
  $x_{20}$ & Bax-Bcl-xL & Bax bound to Bcl-xL & 0\\
  $x_{21}$ & Bad-Scaffold 14-3-3 & Bad bound to Scaffold 14-3-3 & 0\\
  $x_{22}$ & Casp-act & Active caspase & 0\\
  $x_{23}$ & XRCC1(Glu$\sim$Par) & PARP substrate XRCC1 PARylated  & 0\\
 \hline
\end{tabular}
\caption{Mapping between symbol and species name in the ODE model in Equation~\ref{eqn:ode_model_full}.}\label{table:ode_species}
\end{table}

\begin{table}[H]
\centering
\begin{tabular}{||c c l||} 
 \hline
 Symbol & Species & Nominal Value\\ [0.5ex] 
 \hline\hline
  $p_1$ & $s_1$ & 0.01\\
  $p_2$ & $s_2$ & 0.03\\
  $p_3$ & $s_3$ & 20 \\
  $p_4$ & $s_4$ & 0.2 \\
  $p_5$ & $d_1$ & 0.001 \\
  $p_6$ & $d_1$ & 0.0001 \\
  $p_7$ & $d_1$ & 0.002 \\
  $p_8$ & $M$ & 10,000\\
  $p_9$ & $b_1$ & \num{3e-5}\\
  $p_{10}$ & $b_2$ & 0.003 \\
  $p_{11}$ & $b_3$ & 0.003 \\
  $p_{12}$ & $b_4$ & \num{3e-5} \\
  $p_{13}$ & $u_1$ & 0.0001 \\
  $p_{14}$ & $u_2$ & 0.0001\\
  $p_{15}$ & $u_3$ & 0.0001\\
  $p_{16}$ & $u_4$ & 0.0001\\
  $p_{17}$ & $p_1$ & \num{3e-10}\\
  $p_{18}$ & $q_1$ & \num{3e-5}\\
  $p_{19}$ & $a_1$ & \num{2e-10}\\
  $p_{20}$ & $a_2$ & \num{1e-12}\\
  $p_{21}$ & $kf_1$ & 0.001\\
  $p_{22}$ & $kr_1$ & 3.79\\
  $p_{23}$ & $IC_{50}$ & N/A\\
  $p_{24}$ & $kf_2$ & 0.001\\
  $p_{25}$ & $kr_2$ & 0.00379\\
  $p_{26}$ & $kcat_1$ & 1\\
  $p_{27}$ & $kcat_2$ & 1\\
 \hline
\end{tabular}
\caption{Mapping between symbol and parameter name in the ODE model in Equation~\ref{eqn:ode_model_full}. There is no nominal value specified for $p_{23}$=\ic because it is a decision variable and its value is subject to change within our studies.}\label{table:ode_params}
\end{table}

\end{document}